\documentclass [structabstract]{aa}
\usepackage{txfonts}
\usepackage{natbib}
\usepackage{graphicx}
\usepackage{color}
\usepackage{aalongtable}
\newcommand{\ft}[1]{{$^\mathrm{#1}$}}
\bibpunct{(}{)}{;}{a}{}{,}

\begin{document}

\title{Lucky Imaging survey for southern M dwarf binaries
	\thanks{Based on observations made with ESO Telescopes at La Silla or Paranal Observatories under programme ID 082.C-0053}
}

\author{C. Bergfors\inst{1}\thanks{Member of the International Max Planck Research School for Astronomy and Cosmic Physics at the University of Heidelberg}
	\and W. Brandner\inst{1}
	  \and M. Janson\inst{2}
	    \and S. Daemgen\inst{3}
	    \and K. Geissler\inst{1}
	    \and T. Henning\inst{1}
	    \and \\S. Hippler\inst{1}
	    \and F. Hormuth\inst{1}
	    \and V. Joergens\inst{1,4}
	    \and R. K\"ohler\inst{1,5}
}
	
\institute{Max-Planck-Institut f\"ur Astronomie, K\"onigstuhl 17, 69117
  Heidelberg, Germany \\
  \email{bergfors@mpia.de}
  \and University of Toronto, Dept. of Astronomy, 50 St George Street, Toronto, ON, M5S 3H8, Canada
  \and European Southern Observatory, Karl-Schwarzschild Strasse 2, 85748 Garching, Germany 
  \and Zentrum f\"ur Astronomie Heidelberg, Institut f\"ur Theoretische Astrophysik, Albert-Ueberle-Str. 2, 69120 Heidelberg, Germany
  \and Landessternwarte, Zentrum f\"ur Astronomie der Universit\"at Heidelberg, K\"onigstuhl, 69117 Heidelberg, Germany
}

\date{Received 21 January 2010/
			Accepted 1 June 2010}

\abstract{While M dwarfs are the most abundant stars in the Milky Way, there is still
large uncertainty about their basic physical properties (mass, luminosity,
radius, etc.) as well as their formation environment. Precise knowledge of multiplicity characteristics and how they change in this transitional mass
region, between Sun-like stars on the one side and very low mass stars and brown dwarfs on the other, provide constraints on low mass star and brown dwarf formation.}
{In the largest M dwarf binary survey to date, we search for companions to active, and thus preferentially young, M dwarfs in the solar
neighbourhood. We study their binary/multiple properties, such as the multiplicity frequency and distributions of mass-ratio and separation, and
identify short period visual binaries, for which orbital parameters and hence
dynamical mass estimates can be derived in the near future.
}
{The observations are carried out in the SDSS $i^\prime$ and $z^\prime$ band using the Lucky
Imaging camera AstraLux Sur at the ESO 3.5\,m New Technology Telescope. Lucky Imaging
is a very efficient way of observing a large sample of stars at an angular
resolution close to the diffraction limit.
}
{In the first part of the survey, we observed 124 M dwarfs of integrated spectral types M0-M6 and identified 34 new and 17 previously known companions to 44 stars.
We derived relative astrometry and component photometry for these binary and multiple systems.
More than half of the binaries have separations smaller than
1\arcsec and would have been missed in a simply seeing-limited survey. 
Correcting our sample for selection effects yields a multiplicity fraction of 32$\pm$6\%
for 108 M dwarfs within 52\,pc and with angular separations of 0.1$\arcsec$-6.0$\arcsec$, corresponding to projected separations of 3-180\,A.U. at median 
distance 30\,pc. Compared to early-type M dwarfs ($M\ga0.3M_{\sun}$), later-type (and hence lower mass) M dwarf binaries appear to have
closer separations, and more similar masses.
}
{}


\maketitle

\titlerunning{Lucky Imaging survey for southern M dwarf binaries}
\authorrunning{C. Bergfors et al.}

\section{Introduction}

M dwarfs form a link between solar-type stars and brown dwarfs, two mass regions that exhibit very different multiplicity characteristics. Because
properties such as binary fraction, period distribution, and mass-ratio distribution provide important constraints on models of star formation and 
dynamical evolution \citep{Goodwin2007,Burgasser2007}, precise knowledge of multiplicity characteristics and how they change within this transitional mass region is 
important to understanding the formation of low-mass stars and brown dwarfs. 
Repeated astrometric observations of binary systems can also provide dynamical mass estimates, which are crucial to the empirical calibration of the mass-luminosity relation and
evolutionary models.
While being well known for solar-type stars, these relations are not very well 
constrained for lower mass stars. Theoretical models have been shown to underpredict the masses of M dwarfs ($M\la0.5M_{\sun}$) by 5-20\%, and are particularly
inconsistent for masses below $0.3M_{\sun}$ \citep[e.g.,][]{HillenbrandWhite2004}.

It is generally agreed upon that the binary fraction $f_\mathrm{bin}=N_\mathrm{binaries}/N_\mathrm{total}$ 
decreases with decreasing stellar mass \citep[see, e.g., review by][]{Burgasser2007}. While the binary fraction of Sun-like stars is $\approx$57\% 
\citep{DuquennoyMayor1991} over the full range of orbital separations, the fraction of 
multiple stars decreases to $\approx$26-42\% for M0-M6 dwarfs \citep{Delfosse2004, ReidGizis1997, FischerMarcy1992}. For very 
low mass stars ($M<0.1M_{\sun}$) and brown dwarfs, the binary frequency is only 10-30\% \citep[e.g.,][]{Bouy2003,Reid2008,Joergens2008,Goldman2008}. These previous
surveys of M dwarfs are limited to relatively small individual sample sizes, the largest until now being that of \citet{Delfosse2004}, which consisted of 100 stars.

Whether the observed multiplicity characteristics are smooth functions of mass - implying that very low mass stars (VLMSs) and brown dwarfs (BDs) form
like more massive stars - or if another process is primarily responsible for the formation of VLMSs and BDs, is 
debated. The multiplicity distributions of VLMSs and BDs show some important differences from those of Sun-like stars. The semi-major axis distribution
of VLMSs and BDs is narrow and peaks at small separations \citep[3-10\,AU, e.g.,][]{Burgasser2007}, in strong contrast to the separation distribution of solar-type binaries, 
which is wide and peaks at around 30\,AU \citep{DuquennoyMayor1991}. The
mass-ratio distribution also differs for VLMSs and BDs from that of Sun-like stars, showing a clear preference for equal mass binaries \citep[e.g.,][]{Burgasser2007} as 
opposed to the flat 
distribution of the more massive stars \citep{DuquennoyMayor1991}. For M dwarfs, the mass range in-between, \citet{FischerMarcy1992} found a relatively flat mass-ratio
distribution, while \citet{ReidGizis1997} found a preference for almost equal mass systems. 
The differences in binary characteristics have been argued by, e.g., \citet{ThiesKroupa2007}
to support the existence of two populations, 'star-like' and 'BD-like', which are formed by different processes.

The AstraLux M dwarf survey \citep{Hormuth2009} investigates the multiplicity characteristics of low-mass stars using high-resolution Lucky Imaging performed by the two 
AstraLux instruments, AstraLux Norte at the Calar Alto 2.2m telescope \citep{Hormuth2008} and AstraLux Sur at NTT at La Silla \citep{Hippler2009}. The full survey will 
include $\sim$800 stars in the range of spectral types M0-M6 within 52 pc from the Sun, selected from the \citet{Riaz2006} catalogue of young, nearby late-type stars. 
The choice of observing young stars is motivated by the higher sensitivity to substellar companions, which at young ages are still warm and hence brighter and 
easier to detect than around older stars. A 0.072\,$M_{\sun}$ brown dwarf is 3.2 magnitudes brighter in I-band at the age of 0.5\,Gyr than at an age of 5\,Gyr 
\citep{Baraffe2003}. 
Thus, by surveying young M dwarfs we can also detect brown dwarf companions with masses close to the stellar/substellar boundary.
The large sample will allow a detailed statistical analysis of multiplicity characteristics, in the mass region between Sun-like stars and brown dwarfs where these
properties change drastically. 
Follow-up observations of close, nearby multiple systems will also enable dynamical masses to be determined, allowing calibration of the mass-luminosity relation for stars less
massive than 0.5$M_{\sun}$.
We present here the first southern sky sample, consisting of 124 M dwarfs.

\section{Observations and data reduction}
\subsection{Observations}

The first subsample of the 124 nearby M dwarfs presented here (see Table 1) was observed with the AstraLux Sur high resolution camera mounted at the Nasmyth B focus of the ESO 3.5\,m 
New Technology Telescope (NTT) at La Silla on November 12-16, 2008.  The targets were selected from the \citet{Riaz2006} catalogue of $\approx$~1000 nearby active M dwarfs.
All of our targets have spectral types M0-M6 and lie within 52\,pc of the Sun.
We do not have direct age estimates for more than a few individual stars (see Appendix), although
 the \citet{Riaz2006} sample was compiled by correlating 2MASS with ROSAT data, and the sample as a whole, based on its typically strong coronal emission and
low tangential velocity ($<40$\,km s$^{-1}$), is very likely young. 

AstraLux Sur \citep{Hippler2009} is a high-speed electron multiplying camera for Lucky Imaging observations at the NTT. The instrument is an almost identical 
copy of the common user AstraLux Norte camera at the Calar Alto 2.2\,m telescope \citep{Hormuth2008}.  

The Lucky Imaging principle is to minimize atmospheric seeing effects by taking many ($\sim$~10\,000) very short exposures ($\sim$~10\,ms) of the target, thereby effectively 
''freezing'' the
atmosphere in each image. Only the least distorted few percent of the frames, selected on the basis of Strehl ratio, are then combined to achieve almost diffraction-limited resolution. 
The Drizzle algorithm \citep{FruchterHook2002} shifts and adds the slightly undersampled raw images by centering on the brightest pixel, thereby generating an oversampled
output image with a pixel scale of $\approx15.37$\,mas \citep{Hormuth2008}.

On each night of observations, the M dwarf targets were observed in either the SDSS $i^\prime$ or $z^\prime$ filter. Each star was observed in full-frame mode (FoV 15.74\arcsec, integration time 29.45\,ms) and in 
some cases, if the 
flux was high enough, in subframe mode (FoV 7.87\arcsec, integration time 15.29\,ms), allowing for shorter integration times and hence less distortion by atmospheric 
turbulence. Twilight sky-flats were 
obtained whenever the weather conditions were suitable, otherwise we used dome flats.
Astrometric reference stars in 47 Tuc and Trapezium \citep [see][]{Koehler2008} were observed several times each night, allowing us to determine the
platescale and detector orientation. We assume atmospheric refraction to cause a negligible amount of field distortion
($\sim$1\,mas) since separations between the binaries are small. The IRAF \textit{geomap} procedure was used to determine the 
platescale of the drizzled images to be 15.373\,mas/px with a mean scaling uncertainty of 0.002\,mas/px, and a rotation angle of $1.71\degr\pm0.3\degr$.

\subsection{Photometry and astrometry of the candidate binaries/multiples}
Binary separations, position angles, and magnitude differences in SDSS $i^\prime$ and $z^\prime$ filters were obtained for each binary/multiple system by fitting model PSFs from a set 
of reference stars \citep[see][]{Bouy2003}. We used single stars from our observed sample with symmetric PSFs as references. The astrometric and photometric values presented
are weighted averages of several measurements. The weighting is based on the residuals of the PSF fits. In our analysis, we primarily used the 
highest quality 10\% selection of 10\,000 
integrations with 30\,ms exposure time each, yielding a total integration time of 
30\,s per target and filter. In a few rare cases, we used the 1\% selection to achieve a slightly higher astrometric accuracy. 

Since the Lucky Imaging produces a stellar PSF with an almost diffraction-limited core and a seeing halo, high-pass filtering was implemented before fitting the model 
PSF when the stellar companion was much fainter than the primary star and close enough to reside within the halo. For the astrometric parameters (binary separation and
position angle), we used only the $z^\prime$-band images since they are affected by less atmospheric refraction than the $i^\prime$-band frames. For the wide binaries 
(separation$\rho\ga2$\arcsec),
the PSFs of the companions do not overlap and we used the IRAF aperture photometry task \textit{phot} for the astrometry and photometry. This approach produces results with 
approximately the same uncertainty as the PSF fitting procedure. 
A combination of the two procedures was employed in a
few cases for the triple systems. The dominant errors in the determined position angles arise from the uncertainty in field rotation (see Sect. 2.1) and is therefore
assumed to be 0.3\degr~for all systems. The average error in separation is 4\,mas.

\begin{figure*}
 \centering
   \includegraphics[width=15cm]{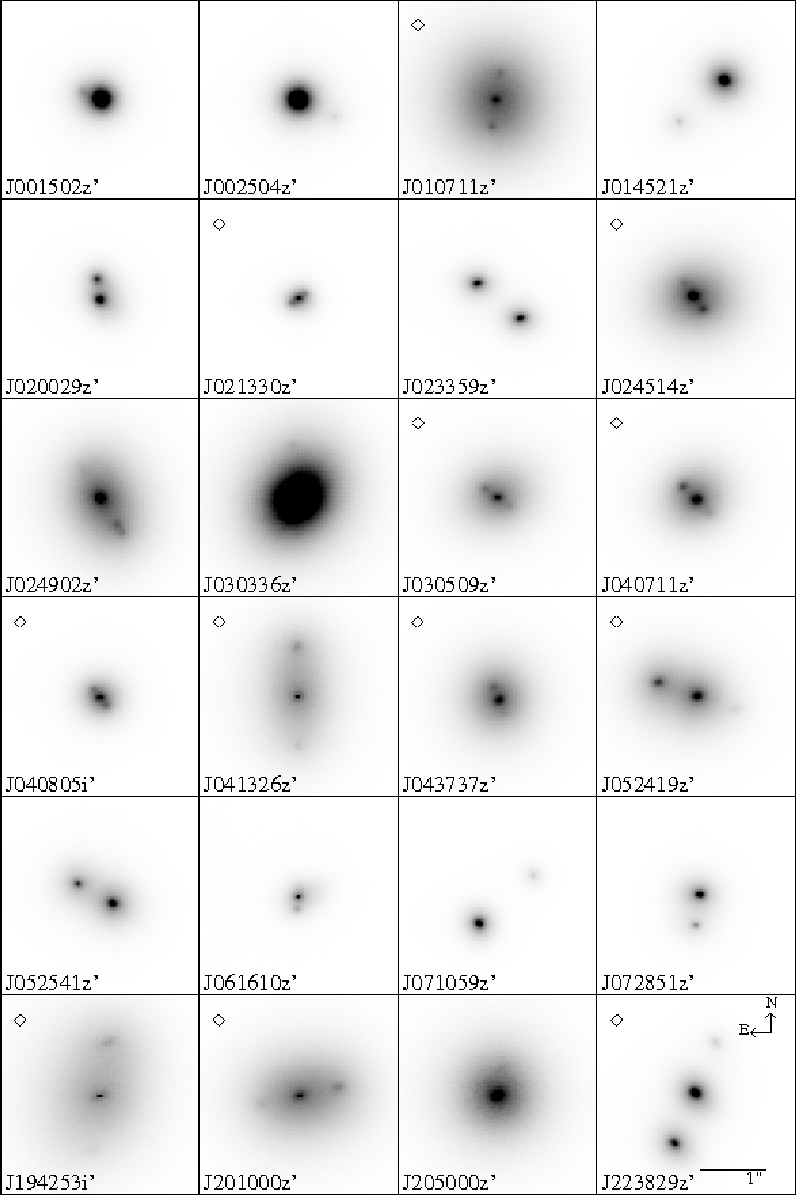}
    \caption{AstraLux Sur images of the systems closer than 1\arcsec. The last character in the ID refers to
the filter in which the star is imaged (SDSS $i^\prime$ or $z^\prime$).The images are shown in a logarithmic intensity scale. 
The scale and orientation is the same for all images and is shown in the bottom right image. The only physical triple system in the figure is J024902. What 
appears as a third star at $180\degr$ angle from the true secondary in some images is an effect of the Lucky Imaging drizzle combination described in Sect. 2.2.
Images affected by this effect are marked with a diamond in the upper left corner.}
 \end{figure*}

\begin{figure*}
 \centering
   \includegraphics[width=15cm]{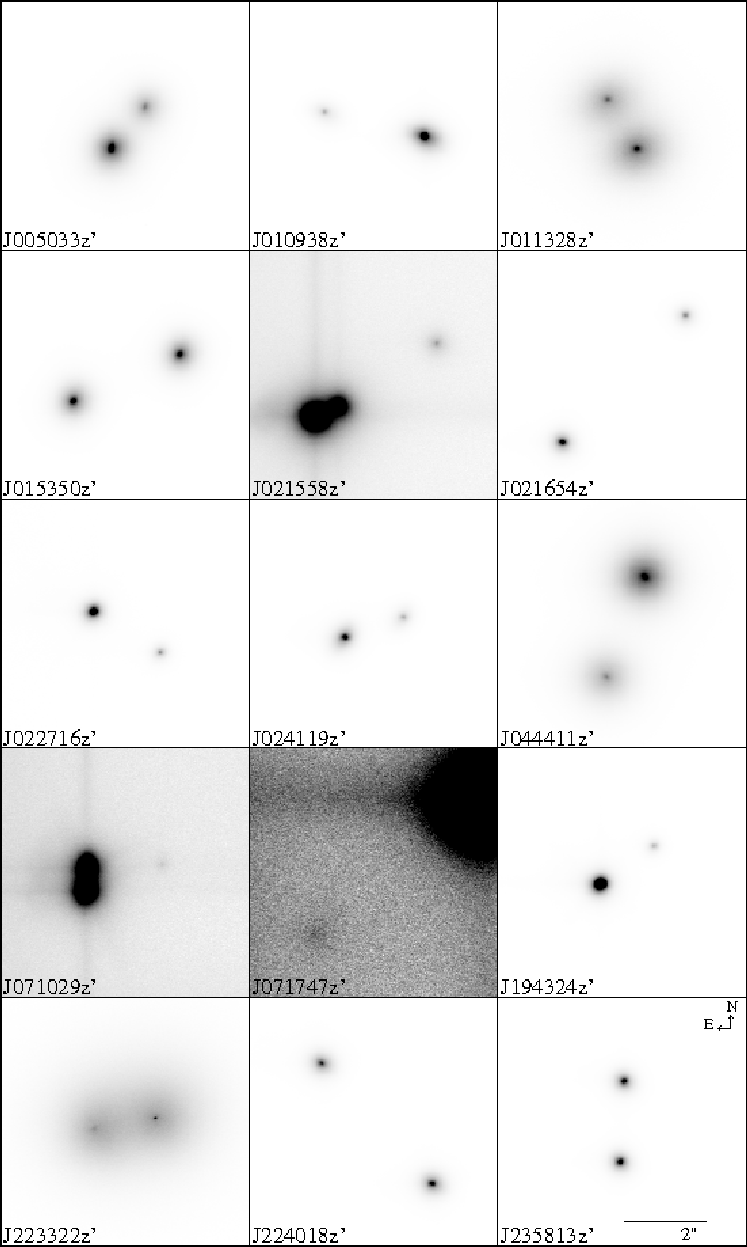}
    \caption{AstraLux Sur images of the systems with separations between 1\arcsec~and 5.5\arcsec. The last character in the ID refers to
the filter in which the star is imaged (SDSS $i^\prime$ or $z^\prime$).The scale and orientation is the same for all images and is shown in the bottom right image.
The images are shown in a logarithmic intensity scale, except for J021558$z^\prime$, J071029$z^\prime$, and J071747$z^\prime$, which are shown on 
a logarithmic square root scale.}
 \end{figure*}

If the two stars are close and of similar magnitude, the Lucky Imaging drizzle combination sometimes centres on the 
secondary star instead of the primary, leading to the appearance of a fake third stellar component in the image. In that case, a ghost stellar image appears at the same 
separation from the primary but at a 180\degr~angle from the real secondary. To recover the flux ratio of the two ''real'' binary components from the fake triple,
we measured the flux of the three components and used the ''de-tripling'' equation of \citet{Law2006} given by

\begin{equation}
   F_\mathrm{R}={2I_{13}\over I_{12}I_{13}+\sqrt{I_{12}^2I_{13}^2-4I_{12}I_{13}}}
\end{equation} 
where $F_\mathrm{R}$ is the true binary flux ratio, $I_{12}= F_1/F_2$ and $I_{13}= F_1/F_3$.

Table 1 lists the complete sample of observed stars with integrated spectral type, distance, J magnitude, and log[L$_\mathrm{x}$/L$_\mathrm{bol}$] from \citet{Riaz2006}, the 
filter(s) in which the star 
was observed and corresponding epoch. Table 2 lists the astrometric and photometric properties derived for the binary/multiple systems in our sample. Component A is 
the primary star, which is 
defined as the brightest of the components in $z^\prime$-band. Figure 1 shows all observed multiple
systems with separations closer than 1\arcsec,~and Fig. 2 shows the wider systems with separations of between 1\arcsec~and 5.5\arcsec. Brightness differences achieved are
typically 3.5 magnitudes for angular separations $\sim0.5\arcsec$ and $\ge6$ magnitudes at $\sim1\arcsec$ (Fig. 3).

\section{Results}
\subsection{Stellar ages and spectral types}
The observed sample is, as a whole, assumed to be young ($\la600$\,Myr), based on its typically strong coronal emission and low tangential velocity \citep{Riaz2006}.
As the velocity dispersion of stars steadily increases with time \citep[e.g., ][]{SeabrokeGilmore2007}, the low tangential velocities of smaller than 40\,km\,s$^\mathrm{-1}$
of the stars in the \citet{Riaz2006} sample combined with activity indicators provide evidence of their youth. \citet{Holmberg2009}, e.g., calibrated the age-velocity relation
(AVR) for FG stars. By scaling their 3D AVR of FG stars to the tangential (2D) velocity dispersion of our sample of M dwarfs, we derive an upper age limit of $\approx$1\,Gyr.

The spectral types of the individual components in the multiple systems were estimated to a precision of $\pm1$ subclasses following the method of
\citet{Daemgen2007}. 
We assumed that the flux ratios of the individual components obtained from the PSF fitting are linearly related to the integrated spectral types provided by \citet{Riaz2006}.
This relation was combined with the \citet{KrausHillenbrand2007} magnitude - spectral type relations, using linear interpolation to derive individual spectral types in
0.5 subclasses. Table 3 summarises the separate component spectral types. 
The spectral types were determined from observations in both filters $i^\prime$ and $z^\prime$ when available,
which are in most cases consistent and otherwise noted in the Appendix. For some stars, we derived primary star spectral types that are 0.5 subclasses 
earlier than the integrated spectral types. The primary spectral type range for the multiple systems is thus K7.5-M5.5 (see Table 3). Only the systems where the
primary star has a spectral type M0 or later are used in the statistical analysis.

The \citet{KrausHillenbrand2007} relations can be used for spectral types no later than L0. However, we estimate that five of the companions are of later spectral type. 
Four of these objects were only observed in $z^\prime$ filter (see Appendix). For 
these five companions, we do not determine the spectral types in any more detail than ``later than L0'' 
until we can assign more precise spectral types using future spectroscopic observations. The
multiple systems containing these faint objects are excluded from the following mass ratio analysis, because of their unknown spectral types (and hence unknown masses). 

\subsection{Binary/multiplicity fraction}
In our sample of 124 observed M dwarfs in the integrated spectral type range M0-M6, we find 51 companions belonging to 44 stars in the angular separation 
range 0.1\arcsec-9.5\arcsec~and 
$z^\prime$-band magnitude difference $0<\Delta z^\prime<6.9$. The observed number of single:binary:triple:quadruple stars is 80:38:5:1. However, the survey is most 
likely insensitive to companions fainter than $\Delta z^\prime\ga2$ in the 
angular separation range 0.1\arcsec-0.5\arcsec,~
and is incomplete for separations greater than 6\arcsec~because of the small FoV. Figure 3 shows the $z^\prime$-band magnitude difference achieved 
as a function of the component angular separation and the typical $5\sigma$ detection limit. 
Figure 4 depicts the number of binaries per angular separation. The distribution is strongly peaked at close separations, with more than half of the 
companions being within 1\arcsec~to the primary star,
suggesting that the vast majority of the observed binaries are indeed physical companions and not the product of background star contamination. 
While most companions were discovered in this survey (34 stars, see Table 2), some of the binaries in our sample were already known to be comoving pairs and some are confirmed 
here by second epoch observations (17 companions, see Table 2 and Appendix).

\begin{figure}
 \resizebox{\hsize}{!}{\includegraphics{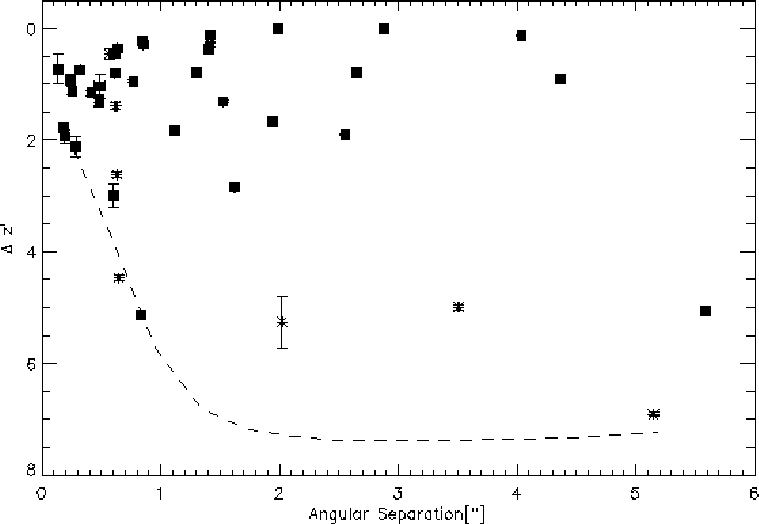}}
  \caption{Observed $z^\prime$-band magnitude difference $\Delta z^\prime$ as a function of angular separation. Squares denote binary systems and asterisks components 
in triple/quadruple systems. Only systems selected from the criteria in Sect. 3.2 are included. The dashed line corresponds to the typical $5\sigma$ detection 
limit in these observations.}
\end{figure}

\begin{figure}
 \resizebox{\hsize}{!}{\includegraphics{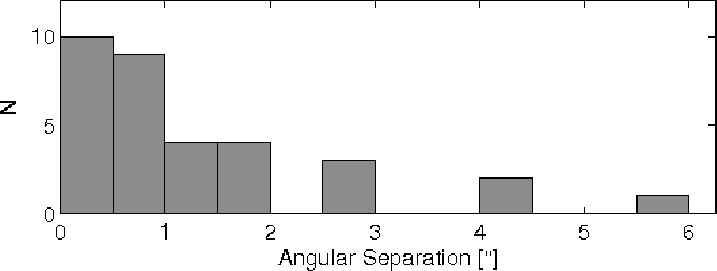}}
    \caption{Binary separation in arcsec for all observed binary systems with separation $\rho\le6.0\arcsec$. The triple and quadruple systems are not included. Note that more than half of the binaries are closer than 1\arcsec, indicating
that the vast majority of the binaries in the sample are physical companions and not background stars. }
 \end{figure}

For the following statistical analysis (multiplicity fraction, mass ratio distribution, and separation distribution), we exclude stars/systems
\begin{itemize}
 \item that lie farther than 52\,pc from the Sun (J06061342-0337082);
  \item components of the binary/multiple systems with separations greater than 6\arcsec~from the primary star (the components J06583980-2021526C, J08224744-5726530C,
and the systems J21103147-2710578 and J22171899-0848122);
  \item binary/multiple systems for which we derive primary spectral type earlier than M0 (J01452133-3957204, J04071148-2918342, J04373746-0229282), or which are part of 
a wider known system containing a primary star of spectral type earlier than M0 (J04373746-0229282, J07174710-2558554);
  \item 'single' stars that are not really single but part of a wide, known system partly outside our field of view.
\end{itemize}
This ensures that
\begin{itemize}
 \item all the single stars in our sample are indeed single, to the best of our knowledge;
  \item the binary/multiple statistics is limited to stars/systems with primary spectral type M0-M6, for stars that lie within 52\,pc of the Sun and have 
separations in the range $0.1\arcsec\leq\rho\leq6.0\arcsec$ (see Table 1).
\end{itemize}

The observed multiplicity frequency $f_\mathrm{obs}=N_\mathrm{Multiple}/N_\mathrm{Total}$ is, after this selection, $35\pm6\%$ (Poisson errors), where $N_\mathrm{Multiple}$ 
is the number of 
binary or multiple systems (38) and $N_\mathrm{Total}$ is the number of observed systems (108).
Figure 5 shows the observed multiplicity fraction for each primary spectral type. The multiple systems included in the following analysis can be found in Table 2, and
the number of single:binary:triple:quadruple systems in Table 4.

To compute the actual multiplicity frequency, we need to consider two effects: (i) at small separations, we detect more equal brightness binaries than systems with large 
component brightness differences, and (ii) a brightness-limited sample is biased in favour of (previously unresolved) binaries or multiple systems compared to single stars. 

Assuming that the flux ratio distribution is independent of the separation in the observed range (which can be transformed into a flat mass ratio distribution), 
we estimate the number of multiple systems of close separations that we miss using the following method.
We divide the number of binaries in Fig. 3 of observed $\Delta z^\prime$ as a function of angular separation $\rho$ into four different regions of interest. Assuming that 
our sample is complete to $\Delta z^\prime\la5.5$ between angular separation 0.5\arcsec-3\arcsec~and complete to $\Delta z^\prime\la2.5$ for closer separations, 
the ratio of companions in the region
$\rho=0.5\arcsec-3\arcsec$, $\Delta z^\prime=2.5-5.5$ and 
$\rho=0.5\arcsec-3\arcsec$, $\Delta z^\prime=0-2.5$ is the same as the ratio of companions in 
$\rho=0.1\arcsec-0.5\arcsec$, $\Delta z^\prime=2.5-5.5$ and
$\rho=0.1\arcsec-0.5\arcsec$, $\Delta z^\prime=0-2.5$. 
This would result in the survey missing two binary companions in the close separation - high flux ratio region, hence the total multiple fraction should be increased 
to $37\pm6\%$.

We compute the multiplicity fraction for a volume-limited sample, $f^\prime$, following the method and Eq. (4) of \citet{Burgasser2003}

\begin{equation}\label{eq*}
   f^\prime = \frac{f^\prime_\mathrm{obs}}{f^\prime_\mathrm{obs}+\alpha(1-f^\prime_\mathrm{obs})}
\end{equation}
where $f^\prime_\mathrm{obs}=0.37$ is the fraction of observed binaries after sensitivity correction.
\citet{Burgasser2003} consider $\alpha$ values in the range 2.8, corresponding to only equal brightness systems, to 1.9, which corresponds to a flat flux
ratio distribution. The distribution of $z^\prime$-band brightness ratios (see Table 2) in our sample is more peaked towards unequal systems (on a linear brightness ratio 
scale), resulting in $\alpha = 1.73$. According to Eq. (4) of \citet{Burgasser2003}, this then yields a multiplicity fraction for a volume-limited sample 
of $f^\prime=25\pm6\%$. 

However, the \citet{Riaz2006} sample is based on a correlation of M dwarf candidates selected from the 400 million sources in the 2MASS point source catalogue 
\citep[PSC, angular resolution $\sim2\arcsec$,][]{Cutri2003} with the 
150\,000 sources in the ROSAT All Sky Survey \citep[RASS, angular resolution $\sim30\arcsec$, ][]{Voges1999}, thus the brightness limit is imposed by the X-ray 
luminosity of the sources. Hence, we need to correct for the excess
of multiple systems as two or more stellar components emit more X-rays than the corresponding primary component would do if it were single. We can do this straightforwardly by simply
examining all our a posteriori known multiple systems and determining which ones would not have been included in the sample if the primary had been single. X-ray counts and
errors are available from ROSAT \citep{Voges1999} for each of the 44 multiple systems (except for one system, J20500010-1154092, which is counted as a non-detection here).
Given that the components in any given system should be coeval, it is assumed that the X-ray brightness depends only on the stellar luminosity. According to \citet{Riaz2006},
$L_{\rm X} / L_{\rm bol}$ is roughly constant as function of spectral type, hence to a reasonable approximation the X-ray count rate can be assumed to be directly proportional
to the brightness fraction in $z^\prime$-band in linear units. Thus, we use the known $\Delta z^\prime$ for each system in combination with the unresolved X-ray count rate 
to estimate the 
rate for the primary component alone. If the new value results in $S/N < 3.3$, the multiple system in question is counted as having been positively selected for and is 
excluded for the purpose of calculating the multiple fraction for a volume-limited sample, where $S/N = 3.3$ is the relevant criterion for detection according to the tables of 
\citet{Voges1999}.
In total, 7 systems are identified as contaminants in this way. Hence, applying corrections for the X-ray flux limit as described above, it follows that the 
multiplicity fraction $f$ is given by $f = (38-7) / (108-7)*1.053 = 32\pm6$\%.

While both multiplicity fractions $f$ and $f^\prime$ agree within the uncertainties, in the following we assume a multiplicity fraction $f=32\pm6\%$, 
as the brightness limit is primarily imposed by the X-ray luminosity. We note that some overabundance of short-period binaries (P$<$20 days) might be present in 
the X-ray selected sample, but this cannot be quantified until future radial velocity observations have been performed. We also note that this fraction might still include a small contamination by non-physical
(``optical'') binaries, as second-epoch observations for some of the systems are still pending, although we reiterate that the fraction of binaries that are merely optical
must be very small (see Fig. 4).

\begin{figure}
 \resizebox{\hsize}{!}{\includegraphics{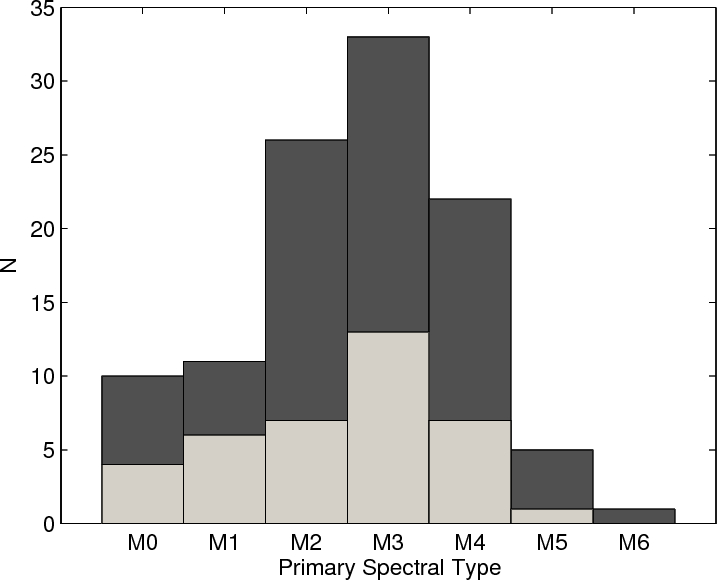}}
    \caption{Multiplicity fraction for each spectral type. The dark grey bins show all observed stars of each spectral type after selection criteria described
in Sect. 3.2 have been imposed on the observed sample. The light grey bins represent the number of
multiple systems in the survey. }
 \end{figure}

\subsection{Mass ratio distribution}
The individual component photometric spectral types from 
Sect. 3.1 are transformed to approximate masses using the mass estimates of \citet{KrausHillenbrand2007} for young ($\sim500$\,Myr) stars. 
We interpolate linearly
to obtain masses for subclasses of 0.5 and calculate the mass-ratios, $q=M_2/M_1$.
The binaries where the secondary star is suspected to be an L dwarf are not included in the mass-ratio distribution because of the high uncertainties in mass. We
also exclude components at separations greater than $6\arcsec$ from the primary star and systems where the primary star is of spectral type earlier than M0 (see Tables 2
and 3). 
Since we also wish to include the triple systems in the multiplicity statistics, and all of our triple systems consist of one close pair and 
one wider component, we follow \citet{ReidGizis1997} and calculate the triple mass-ratio as if the system consists of two separate binary 
systems, one close pair $q_\mathrm{close}=M_\mathrm{B}/M_\mathrm{A}$ and one wider system with the combined mass of the close system as the higher mass component, e.g., 
$q_\mathrm{wide}=M_\mathrm{C}/(M_\mathrm{A}+M_\mathrm{B})$.
The quadruple system J06583980-2021526 contains one close pair of spectral types M4+M4 and two more distant suspected L dwarfs, (one of which is also outside the
$6\arcsec$ limit).
This system is, therefore, treated as a regular binary system, ignoring the two fainter components. 

The mass-ratio distribution has been seen to vary from a flat distribution among solar-type stars to peak at almost equal mass systems for VLMSs and brown 
dwarfs \citep[see e.g.,][and references therein]{Allen2007}. Figure 6 shows the mass-ratio distribution for our M0-M5.5 binaries compared 
to the distribution for all known
VLMS and brown dwarf binaries compiled from the Very Low Mass Binaries Archive\footnote{http://www.vlmbinaries.org} (total system mass $<0.2M_{\sun}$). We applied small updates to 
the July 28, 2009 version of
the archive. Almost equal mass binaries are preferred for VLMSs/brown dwarfs, but the M dwarf distribution is much flatter.
While our sensitivity limit makes our survey incomplete at the low mass-ratio end of the distribution, equal mass systems should easily be seen.
The lack of a peak near $q\sim1$ is therefore a real property of the M dwarf binary systems in the separation range $0.1\arcsec-6.0\arcsec$. 
We note that the mass-ratio distribution for VLMSs and BDs might be flatter in the case of very young systems \citep{Burgasser2007}. The samples are however very
small, even if we account for more recently discovered systems, and we are therefore unable to address the age effects. No correlation between 
mass-ratio and component separation is seen in our sample.

\begin{figure}
 \resizebox{\hsize}{!}{\includegraphics{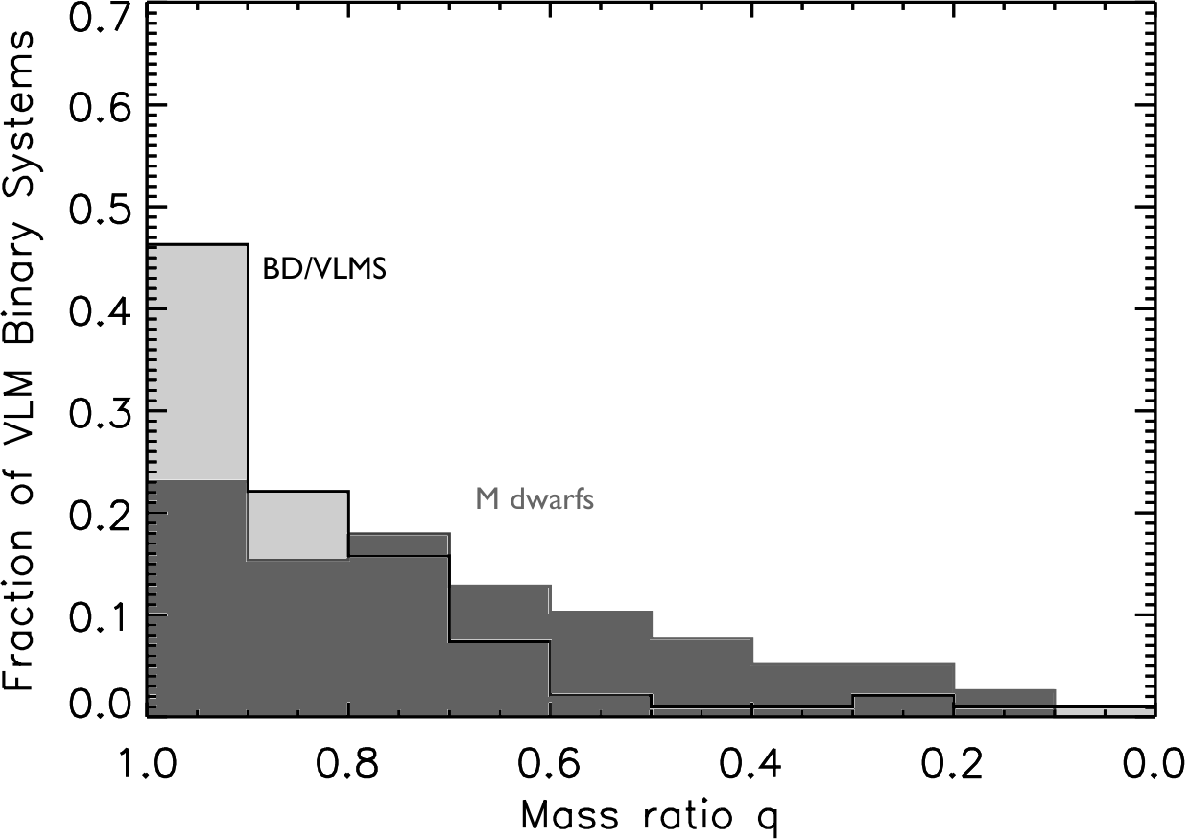}}
    \caption{Mass-ratio distribution. The dark grey distribution shows the mass-ratios of the AstraLux M dwarf binaries. The light grey distribution shows all known VLM 
binaries ($M<0.2M_{\sun}$) from the Very Low Mass Binaries Archive at http://www.vlmbinaries.org for comparison. Triple systems are included as two binaries as described in 
Sect. 3.3. Systems where one component is a suspected L dwarf, where the primary star is of spectral type earlier than M0, and companions with greater separation
than $6\arcsec$ from the primary are not included.}
 \end{figure}

When we divide our sample into early M dwarfs of primary spectral type M0-M3 ($M\ga0.3M_{\sun}$) and late M dwarfs of spectral type M3.5-M5.5 ($M\la0.3M_{\sun}$), 
we see some indication of a 
peak in the distribution around $q\ga0.7-0.8$ for the late type M dwarfs that is not present in the very flat $f(q)$ distribution of the early type M dwarfs (see Fig. 7). 
Assuming that our survey is not complete for mass-ratios $q<0.4$, a Kolmogorov-Smirnov (K-S) test shows that the probability that the 'early-M' and the 'late-M' mass-ratios 
are drawn from the same distribution is 10\%.
This might indicate that the shape 
of the mass ratio distribution is a function of mass, which approaches the $q\sim1$ peak for the lower mass stars.
However, this division into seemingly different populations should be assumed with caution. For mid- to late-M dwarfs, the mass - spectral type relation becomes very steep. 
Thus, a large brightness difference corresponds to only a very small change in 
mass for lower mass objects. Hence, a detection limit of $\Delta z^\prime\la1.5$ magnitudes, which we assume to be valid for all stars in the sample, corresponds to a 
mass-ratio
completeness $q\ga0.4$ for early-type M dwarfs, while the same detection limits correspond to completeness only for $q\ga0.6$ for an M3.5 primary star and $q\ga0.8$ for an M5
primary. While the missing $q\sim1$ peak is an unbiased feature, the sensitivity to lower $q$ values is strongly dependent on spectral type.
With more observations from the full
AstraLux M dwarf survey, we will be able to investigate these distributions in greater detail.

\begin{figure}
 \resizebox{\hsize}{!}{\includegraphics{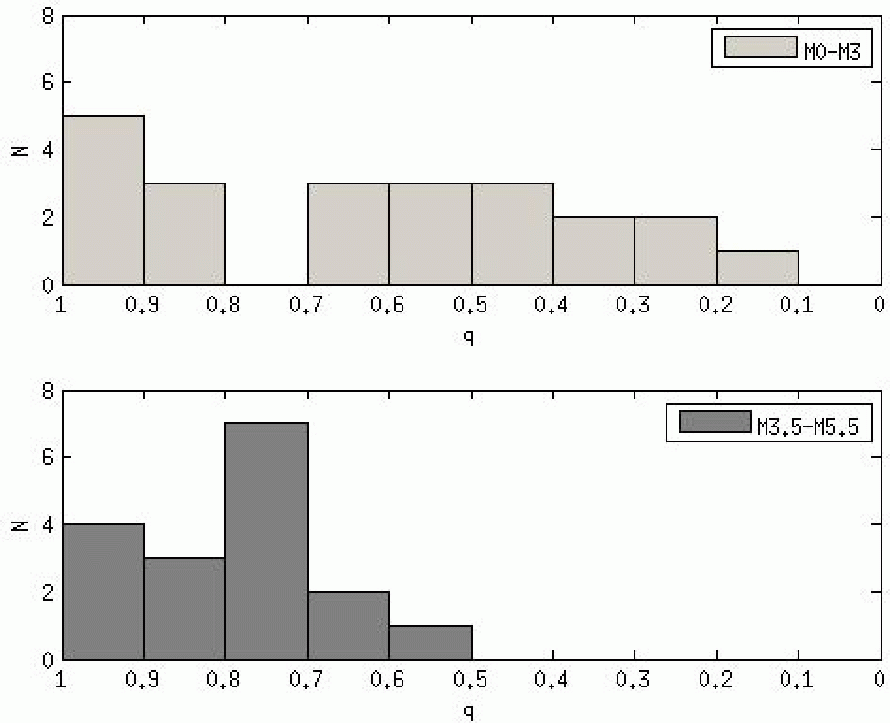}}
    \caption{Mass-ratio distribution divided into early-M type primaries (M0-M3) and late-M type (M3.5-M5.5). 
Triple systems are included as described in Sect. 3.3. Systems where one component is a suspected L dwarf and components at separations greater than $6\arcsec$ are
not included.}
 \end{figure}

\subsection{Distribution of separations}
From the parallax distances, if available, and otherwise the spectroscopic distances provided by \citet{Riaz2006}, we calculate the projected separation in 
astronomical units. The uncertainty in spectroscopic distance according to \citet{Riaz2006} is $37\%$.
Figure 8 shows the distribution of projected separation of all binaries and triples in our M dwarf sample compared to that of all known VLMS/BD binaries from the Very Low Mass Binaries 
Archive. 

\begin{figure}
 \resizebox{\hsize}{!}{\includegraphics{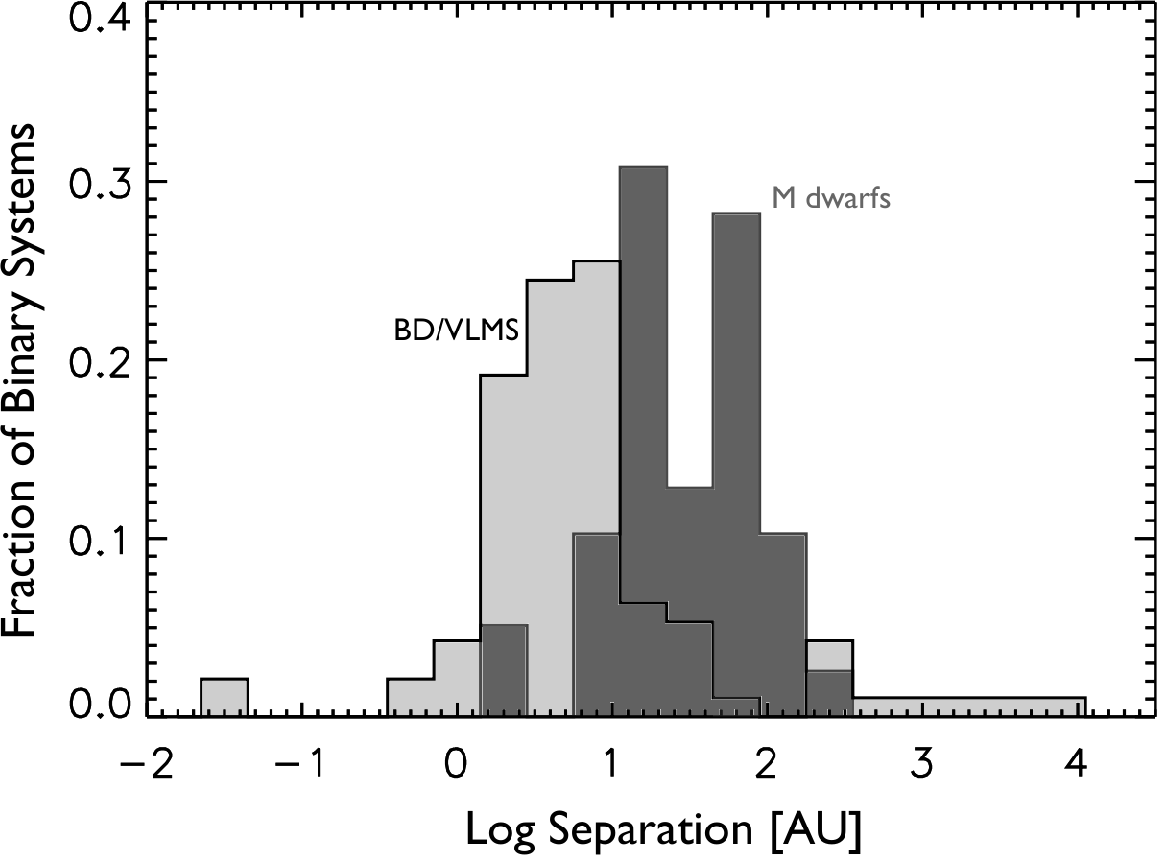}}
    \caption{Projected separation distributions. The darker grey shows the M dwarf binaries/triples in our sample and the light grey shows VLMS/BD binaries from the 
Very Low Mass Binaries Archive.}
 \end{figure}

As for the mass-ratio distribution, we divide the observed systems into two groups, containing approximately equal number of systems, to see if the 
separation distribution is the same for 'early M' and 'late M' type binaries divided at $M\approx0.3M_{\sun}$. Figure 9 shows the respective mean semi-major axis 
distributions, where the projected separation has been multiplied with 1.26 to account for random orbital elements \citep{FischerMarcy1992}.
We performed a K-S test, which yielded a 9\% probability that 
the distributions are alike. We note that the distributions may peak at close systems in the 'late M' subsample, however more data is necessary 
to determine whether this is a real property or not. 

\begin{figure}
 \resizebox{\hsize}{!}{\includegraphics{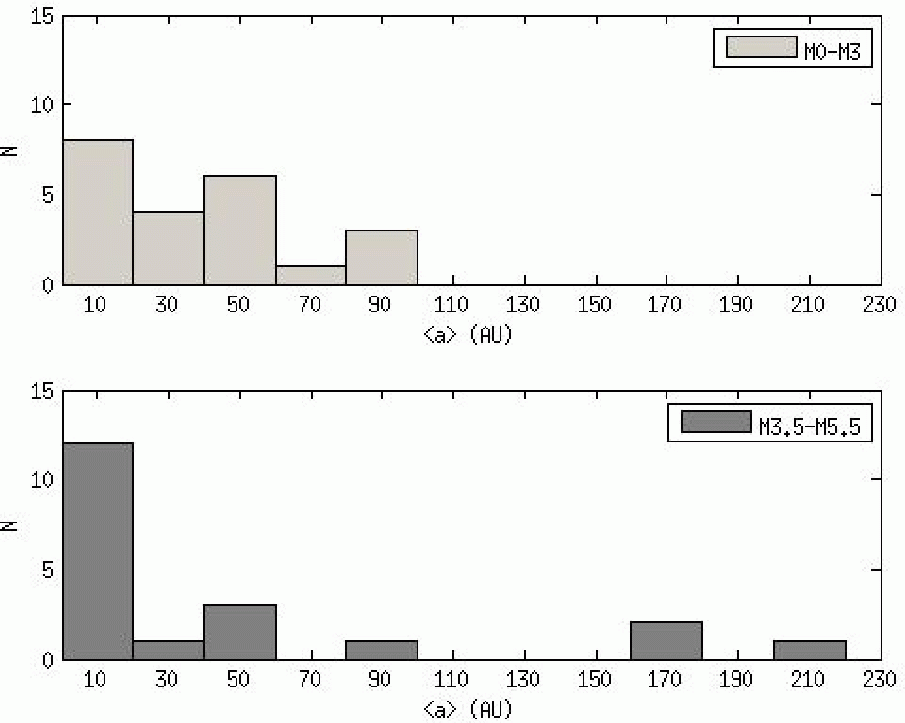}}
    \caption{Distribution of mean semi-major axis for 'early M' and 'late M' primary spectral types. The binaries are divided into two groups: `early type M', consisting of the 
binaries with a primary spectral type M0-M3,
 and 'late type M' for primary spectral types M3.5-M5.5. All separations are measured from the primary stars.}
 \end{figure}

\section{Discussion}
M dwarfs comprise a transitional region within which the multiplicity properties change from being similar to those of solar-type stars to the very different characteristics
of very low mass stars and brown dwarfs. Smaller surveys of different mass ranges have provided some insight into the transitional behaviour.

We observed 124 nearby M dwarfs from the \citet{Riaz2006} catalogue. Forty-four of our targets were observed to have potential binary/multiple companions within 
$0.1\arcsec-9.5\arcsec$ of the primary star. Most of these companions were previously unknown.

We have estimated the multiplicity fraction for M0-M6
young ($\la600$\,Myr) dwarfs with angular separations $0.1\arcsec-6\arcsec$, corresponding to projected separation 3-180\,A.U. at median distance 30\,pc, 
in this largest sample to date to be $32\pm6\%$. 
While differences in the binary fraction have been found in various nearby star-forming regions \citep{Leinert1993, Ghez1993, BrandnerKoehler1998, Koehler2006}, 
observations of the 90\,Myr old $\alpha$ Persei and the 600\,Myr old Praesepe clusters suggest that the companion star fraction does not significantly decline over an age
range from 90\,Myr to 5\,Gyr \citep{Patience2002}. We therefore did not expect there to be a strong evolution in binary properties from ages $\sim100$\,Myr to a few Gyr. 
Our derived
$f_\mathrm{Mult}$ is consistent with previous
surveys in the same mass and separation range for field dwarfs \citep[e.g., the multiplicity fraction of the \citet{FischerMarcy1992} sample for
M0-M4 systems with linear separation 2.6\,AU$<a<$300\,AU is $f_\mathrm{bin}=28\pm9\%$, according to ][]{Close2003} and young M dwarfs \citep[23\% for $<300$\,Myr
M0-M5 binaries with separations 1.6-300\,AU, ][]{Daemgen2007}. This is a higher $f_\mathrm{Mult}$ than found in high resolution imaging surveys
of later type M field dwarfs (e.g., \citet{Law2008} 13.6\% for M4.5-M6; \citet{Close2003} 15\% for M8-L0.5; \citet{Siegler2005} 9\% for M6-M7.5).

Previous large investigations inferred different behaviours for the mass ratio distribution of M dwarfs. For instance, 
\citet{FischerMarcy1992} found that the mass ratio distribution for M0-M6 dwarfs is relatively flat for the full orbital separation range, while
\citet{ReidGizis1997} found a strong peak for $q\ga0.8$. However, \citet{Delfosse2004} showed in their large survey that the 
distribution of mass-ratio is relatively flat for orbital periods P$>$50 days, while shorter period binaries tend to have equal masses. 

When considering only systems in the survey of \citet{ReidGizis1997}
with M dwarf primaries (M0.5-M5.5) and mean semi-major axis 3.7\,AU$<a<$227\,AU, the same region probed in this survey, we find that the distribution is flat (although the sample is very small 
with only 16 companions). We therefore expect our distribution to be flat as well, in accordance with those of other surveys.
We have found that the mass-ratio distribution of all binaries in the primary spectral type range M0-M5.5 is flatter than the distribution for 
VLMSs and BDs and does not exhibit the prominent peak at $q\sim 1$ detected for the VLMS/BD sample. This is consistent with the results of \citet{ReidGizis1997}
over the same range of linear separations, and a real feature for M dwarfs in the 
observed separation range not affected by observational incompleteness.
To investigate the transitional properties, we divided our sample into two groups of 'early M' and 'late M' primary stars and found that the 'early M'
distribution is relatively flat, while there might be a preference for more similar mass binaries in the 'late M' group. Future analysis
 of the full AstraLux M dwarf survey will allow us to investigate this possible trend in more detail.

\citet{ThiesKroupa2007} argued that the differences in binary characteristics of stars and brown dwarfs point to different but related formation processes for these
two populations. In their survey of M4.5-M6 binaries, \citet{Law2008} found a bimodal separation distribution where the later type M dwarfs peak at close
separations as seen for brown dwarfs, but some of the earlier systems have projected separations greater than 10\,AU. This is indicative of a change in the separation 
distributions at spectral type $\sim$M5, consistent with the \citet{ThiesKroupa2007} predictions of two separate but overlapping populations.
For later type M dwarfs, \citet{Siegler2005} find no binaries with separations $a>10$\,AU in their high-resolution imaging survey of M6-M7.5 field 
dwarfs covering separations 3-300\,AU, and \citet{Close2003} find no binaries wider than 16\,AU for the M8-L0.5 dwarfs in the same separation range.
We divide our separation distribution into 'early M' and 'late M' groups at $M\sim0.3M_{\sun}$. Even though our sample contains
somewhat more massive stars than that of \citet{Law2008}, the distribution of 'early M' multiples is flatter 
than the 'late M' distribution in which more than half of the companions reside within 20\,AU of the primary star. While our results are still subject to relatively large 
statistical uncertainties,
we note that they may indicate that a bimodal distribution also exist for larger separations and higher masses than the 
sample of \citet{Law2008}. This will be investigated further in the larger sample of the complete survey.

These results show that Lucky Imaging with AstraLux Sur is very efficient at detecting binary stars with small angular separations. Several of our newly discovered
companions presented here have been found to be close to the diffraction limit ($\sim0.1\arcsec$). Future follow-up observations with AstraLux Sur will allow us to determine 
orbital motions and hence dynamical masses for the closest nearby systems.

\begin{acknowledgements}
We thank the anonymous referee for helpful comments and suggestions leading to an improved paper.
This publication has made use of the Very-Low-Mass Binaries Archive housed at http://www.vlmbinaries.org and maintained by Nick Siegler, Chris Gelino, and Adam Burgasser.
This research has made use of the SIMBAD database, operated at CDS, Strasbourg, France. 
\end{acknowledgements}

\bibliographystyle{aa}
\bibliography{Bergfors_refs}

\longtab{1}{
\begin{longtable}{lllrrrrl}
\caption{Properties of all stars observed 11-16 November 2008.}\\
\hline\hline
2MASS ID & Other name & SpT\tablefootmark{a} & D [pc]\tablefootmark{a} & J mag\tablefootmark{a} & log $[L_\mathrm{x}/L_\mathrm{bol}]$\tablefootmark{a} & Filter & Epoch\tablefootmark{b}\\
\hline
\endfirsthead
\caption{continued.}\\
\hline\hline
2MASS ID & Other name & SpT\tablefootmark{a} & D [pc]\tablefootmark{a} & J mag\tablefootmark{a} &log $[L_\mathrm{x}/L_\mathrm{bol}]$\tablefootmark{a} & Filter & Epoch\tablefootmark{b}\\
\hline
\endhead
\hline
\endfoot
J00150240-7250326\tablefootmark{c} &		&	M1.0	&	38	&	8.62	&	-3.01	&	$i^\prime$, $z^\prime$	&	2008.86	\\
J00155808-1636578 &		&	M4.0	&	9	&	8.74	&	-3.08	&	$z^\prime$	&	2008.87	\\
J00171443-7032021 &		&	M0.5	&	48	&	9.00	&	-3.23	&	$i^\prime$, $z^\prime$	&	2008.87	\\
J00213729-4605331 &	GJ 3029	&	M3.0	&	20\tablefootmark{d}	&	8.33	&	-2.56	&	$z^\prime$	&	2008.87	\\
J00250428-3646176\tablefootmark{c} &		&	M2.5	&	28	&	8.65	&	-3.30	&	$i^\prime$, $z^\prime$	&	2008.86	\\
J00275035-3233238\tablefootmark{g} &		&	M3.5	&	28	&	8.97	&	-2.75	&	$i^\prime$, $z^\prime$	&	2008.86	\\
J00281434-3227556 &		&	M5.0	&	12	&	10.12	&	-2.85	&	$z^\prime$	&	2008.86	\\
J00503319+2449009\tablefootmark{c} &	GJ 3060A	&	M3.5	&	12\tablefootmark{d}	&	7.92	&	-3.06	&	$i^\prime$, $z^\prime$	&	2008.86	\\
J01025097+1856543 &		&	M4.0	&	13	&	9.51	&	-3.24	&	$z^\prime$	&	2008.86	\\
J01071194-1935359\tablefootmark{c} &		&	M1.0	&	30	&	8.15	&	-3.13	&	$i^\prime$, $z^\prime$	&	2008.87	\\
J01093874-0710497\tablefootmark{c} &	HIP 5443	&	M1.5	&	38\tablefootmark{d}	&	7.96	&	-3.70	&	$i^\prime$, $z^\prime$	&	2008.87	\\
J01112343-0525381 &		&	M3.5	&	35	&	9.47	&	-2.86	&	$z^\prime$	&	2008.87	\\
J01132817-3821024\tablefootmark{c} &		&	M0.5	&	37	&	8.49	&	-3.19	&	$i^\prime$, $z^\prime$	&	2008.88	\\
J01225093-2439505 &		&	M3.5	&	45	&	10.08	&	-3.18	&	$z^\prime$	&	2008.88	\\
J01244246-1540454\tablefootmark{h} &	NLTT 4703	&	M1.5	&	27	&	8.11	&	-4.01	&	$z^\prime$	&	2008.88	\\
J01365516-0647379\tablefootmark{c} &	NLTT 5400	&	M4.0	&	37	&	9.70	&	-2.89	&	$z^\prime$	&	2008.86	\\
J01434512-0602400 &		&	M3.5	&	26	&	8.77	&	-3.00	&	$i^\prime$, $z^\prime$	&	2008.86	\\
J01452133-3957204\tablefootmark{c} &	NLTT 5871 	&	M0.0	&	32\tablefootmark{d}	&	8.43	&	-3.81	&	$i^\prime$, $z^\prime$	&	2008.86	\\
J01511997+1324525 &		&	M1.5	&	31	&	8.56	&	-3.21	&	$z^\prime$	&	2008.86	\\
J01535076-1459503\tablefootmark{c} &		&	M3.0	&	18	&	7.94	&	-3.12	&	$i^\prime$, $z^\prime$	&	2008.86	\\
J02001277-0840516 &		&	M2.5	&	30	&	8.77	&	-3.11	&	$i^\prime$, $z^\prime$	&	2008.86	\\
J02002975-0239579\tablefootmark{c} &		&	M3.5	&	48	&	10.07	&	-2.95	&	$i^\prime$, $z^\prime$	&	2008.87	\\
J02014384-1017295 &	NLTT 6782	&	M4.0	&	17	&	10.03	&	-3.34	&	$z^\prime$	&	2008.87	\\
J02070198-4406444 &		&	M5.5	&	21	&	11.36	&	-2.26	&	$z^\prime$	&	2008.87	\\
J02070786-1810077 &		&	M4.0	&	22	&	10.70	&	-2.73	&	$z^\prime$	&	2008.87	\\
J02133021-4654505\tablefootmark{c} &		&	M4.0	&	13	&	9.43	&	-2.99	&	$i^\prime$, $z^\prime$	&	2008.87	\\
J02155892-0929121\tablefootmark{c} &		&	M2.5	&	26	&	8.43	&	-3.11	&	$i^\prime$, $z^\prime$	&	2008.87	\\
J02164119-3059181\tablefootmark{i} &	GJ 3148 A	&	M3.5	&	14\tablefootmark{d}	&	7.99	&	-3.42	&	$z^\prime$	&	2008.87	\\
J02165488-2322133\tablefootmark{c} &		&	M3.5	&	40	&	9.79	&	-2.50	&	$i^\prime$, $z^\prime$	&	2008.87	\\
J02183655+1218579 &		&	M2.0	&	34	&	8.80	&	-3.40	&	$z^\prime$	&	2008.87	\\
J02224418-6022476 &		&	M4.0	&	10	&	8.99	&	-2.58	&	$z^\prime$	&	2008.87	\\
J02271603-2929263\tablefootmark{c} &		&	M3.5	&	51	&	10.34	&	-3.27	&	$i^\prime$, $z^\prime$	&	2008.87	\\
J02303485-1543248 &	NLTT 8185	&	M2.5	&	39	&	9.29	&	-3.31	&	$z^\prime$	&	2008.87	\\
J02335984-1811525\tablefootmark{c} &		&	M3.0	&	50	&	10.09	&	-2.77	&	$i^\prime$, $z^\prime$	&	2008.87	\\
J02365171-5203036 &		&	M2.0	&	28	&	8.42	&	-3.01	&	$z^\prime$	&	2008.87	\\
J02411510-0432177 &	NLTT 8687 	&	M4.0	&	11	&	9.20	&	-3.64	&	$z^\prime$	&	2008.87	\\
J02411909-5725185\tablefootmark{c} &		&	M3.0	&	45	&	9.85	&	-3.15	&	$i^\prime$, $z^\prime$	&	2008.87	\\
J02414730-5259306 &		&	M2.5	&	27	&	8.48	&	-2.65	&	$z^\prime$	&	2008.87	\\
J02451431-4344102\tablefootmark{c} &		&	M4.0	&	7	&	8.06	&	-3.26	&	$i^\prime$, $z^\prime$	&	2008.88	\\
J02485260-3404246 &		&	M4.0	&	12	&	9.31	&	-2.90	&	$z^\prime$	&	2008.88	\\
J02490228-1029220\tablefootmark{c} &		&	M2.0	&	34	&	8.82	&	-3.21	&	$i^\prime$, $z^\prime$	&	2008.88	\\
J02492136-4416063 &		&	M3.0	&	45	&	9.88	&	-3.30	&	$z^\prime$	&	2008.88	\\
J02543316-5108313 &		&	M1.5	&	34	&	8.67	&	-3.33	&	$z^\prime$	&	2008.88	\\
J03033668-2535329\tablefootmark{c} &	NLTT 9775	&	M0.0	&	39\tablefootmark{d}	&	8.00	&	-3.65	&	$i^\prime$, $z^\prime$	&	2008.88	\\
J03050976-3725058\tablefootmark{c} &		&	M2.0	&	45	&	9.54	&	-3.50	&	$i^\prime$, $z^\prime$	&	2008.88	\\
J03100305-2341308 &	NLTT 10115	&	M3.5	&	35	&	9.41	&	-3.29	&	$z^\prime$	&	2008.88	\\
J03152341-2821404 &		&	M3.5	&	49	&	10.25	&	-2.41	&	$z^\prime$	&	2008.86	\\
J03214689-0640242 &	GJ 3218	&	M2.0	&	16\tablefootmark{e}	&	7.86	&	-4.74	&	$z^\prime$	&	2008.86	\\
J03244056-3904227 &		&	M4.0	&	40	&	9.87	&	-2.27	&	$z^\prime$	&	2008.86	\\
J03271433+2723087 &	NLTT 10933	&	M0.5	&	27\tablefootmark{d}	&	8.64	&	-3.44	&	$z^\prime$	&	2008.86	\\
J03432333-0819412 &		&	M2.5	&	49	&	9.86	&	-3.62	&	$z^\prime$	&	2008.88	\\
J03472333-0158195 &	NLTT 11853	&	M2.5	&	16\tablefootmark{d}	&	7.80	&	-3.14	&	$z^\prime$	&	2008.88	\\
J04071148-2918342\tablefootmark{c} &		&	M0.0	&	51	&	9.06	&	-3.16	&	$i^\prime$, $z^\prime$	&	2008.88	\\
J04080543-2731349\tablefootmark{c} &		&	M3.5	&	43	&	9.89	&	-3.14	&	$i^\prime$, $z^\prime$	&	2008.88	\\
J04093930-2648489 &		&	M1.5	&	49	&	9.51	&	-3.11	&	$z^\prime$	&	2008.88	\\
J04132663-0139211\tablefootmark{c} &		&	M4.0	&	12	&	9.38	&	-2.97	&	$i^\prime$, $z^\prime$	&	2008.88	\\
J04141730-0906544 &		&	M3.5	&	37	&	9.63	&	-2.70	&	$z^\prime$	&	2008.88	\\
J04175717-3827038 &		&	M3.5	&	36	&	9.45	&	-2.91	&	$z^\prime$	&	2008.88	\\
J04213904-7233562 &		&	M2.5	&	50	&	9.87	&	-2.97	&	$z^\prime$	&	2008.88	\\
J04240094-5512223 &		&	M2.5	&	48	&	9.80	&	-3.26	&	$z^\prime$	&	2008.88	\\
J04241156-2356365 &		&	M2.5	&	24	&	8.32	&	-3.70	&	$i^\prime$, $z^\prime$	&	2008.88	\\
J04353618-2527347 &	NLTT 13598	&	M3.5	&	20	&	8.24	&	-3.24	&	$i^\prime$, $z^\prime$	&	2008.88	\\
J04365738-1613065 &		&	M3.5	&	30	&	9.12	&	-2.63	&	$i^\prime$, $z^\prime$	&	2008.88	\\
J04373746-0229282\tablefootmark{c} &	GJ 3305	&	M0.0	&	23	&	7.30	&	-2.71	&	$i^\prime$, $z^\prime$	&	2008.88	\\
J04380252-0556132 &	NLTT 13666	&	M4.5	&	12	&	9.73	&	-3.05	&	$z^\prime$	&	2008.88	\\
J04441107-7019247\tablefootmark{c} &	HD 270712	&	M1.5	&	19	&	7.46	&	-3.62	&	$i^\prime$, $z^\prime$	&	2008.88	\\
J04522441-1649219 &	NLTT 14116	&	M3.0	&	18	&	7.74	&	-3.17	&	$i^\prime$, $z^\prime$	&	2008.88	\\
J05082729-2101444 &		&	M5.0	&	11	&	9.72	&	-3.19	&	$i^\prime$, $z^\prime$	&	2008.88	\\
J05241914-1601153\tablefootmark{c} &		&	M4.5	&	8	&	8.67	&	-3.14	&	$i^\prime$, $z^\prime$	&	2008.88	\\
J05254166-0909123\tablefootmark{c} &	NLTT 15049	&	M3.5	&	22	&	8.45	&	-3.18	&	$i^\prime$, $z^\prime$	&	2008.86	\\
J05332802-4257205 &		&	M4.5	&	6	&	8.00	&	-3.13	&	$i^\prime$, $z^\prime$	&	2008.88	\\
J06045215-3433360 &		&	M5.0	&	4	&	7.74	&	-2.95	&	$i^\prime$, $z^\prime$	&	2008.88	\\
J06061342-0337082\tablefootmark{j} &		&	M2.5	&	57	&	10.15	&	-3.03	&	$i^\prime$, $z^\prime$	&	2008.88	\\
J06161032-1320422\tablefootmark{c} &		&	M4.0	&	31	&	11.35	&	-2.32	&	$i^\prime$, $z^\prime$	&	2008.87	\\
J06224133-2737531 &		&	M3.5	&	35	&	9.43	&	-3.21	&	$z^\prime$	&	2008.87	\\
J06253604-4815598 &		&	M2.5	&	34	&	9.10	&	-3.74	&	$z^\prime$	&	2008.87	\\
J06255610-6003273 &		&	M3.5	&	19	&	8.09	&	-2.90	&	$z^\prime$	&	2008.87	\\
J06525392-0524413 &		&	M2.5	&	30	&	8.71	&	-3.17	&	$z^\prime$	&	2008.87	\\
J06583980-2021526\tablefootmark{c} &		&	M4.0	&	32	&	9.40	&	-3.23	&	$i^\prime$, $z^\prime$	&	2008.87	\\
J07020886-0626206 &		&	M2.0	&	52	&	9.80	&	-2.67	&	$z^\prime$	&	2008.87	\\
J07065772-5353463 &		&	M0.0	&	41	&	8.54	&	-3.17	&	$z^\prime$	&	2008.87	\\
J07102991-1637350\tablefootmark{c} &		&	M2.5	&	49	&	9.75	&	-3.02	&	$i^\prime$, $z^\prime$	&	2008.87	\\
J07105990-5632596\tablefootmark{c} &		&	M1.5	&	52	&	9.61	&	-2.43	&	$i^\prime$, $z^\prime$	&	2008.87	\\
J07120447-3048526 &		&	M2.5	&	47	&	9.71	&	-3.44	&	$z^\prime$	&	2008.87	\\
J07174710-2558554\tablefootmark{c} &		&	M2.0	&	47	&	9.53	&	-3.19	&	$z^\prime$	&	2008.88	\\
J07285137-3014490\tablefootmark{c} &	GJ 2060	&	M1.5	&	16\tablefootmark{d}	&	6.62	&	-3.04	&	$i^\prime$, $z^\prime$	&	2008.86	\\
J08224744-5726530\tablefootmark{c} &	LHS 2005	&	M4.5	&	8	&	8.63	&	-3.14	&	$i^\prime$, $z^\prime$	&	2008.88	\\
J19425324-4406278\tablefootmark{c} &		&	M3.5	&	36	&	9.43	&	-3.12	&	$i^\prime$, $z^\prime$	&	2008.87\tablefootmark{m}	\\
J19432464-3722108\tablefootmark{c} &		&	M3.5	&	31	&	9.20	&	-3.31	&	$i^\prime$, $z^\prime$	&	2008.87	\\
J19513587-3510375 &		&	M4.0	&	8	&	8.58	&	-3.23	&	$z^\prime$	&	2008.87	\\
J20100002-2801410\tablefootmark{c} &		&	M3.0	&	26	&	8.65	&	-3.16	&	$i^\prime$, $z^\prime$	&	2008.87	\\
J20194981-5816431 &		&	M6.0	&	12	&	10.66	&	-2.28	&	$i^\prime$, $z^\prime$	&	2008.87	\\
J20500010-1154092\tablefootmark{c} &		&	M3.5	&	38	&	9.68	&	-3.58	&	$i^\prime$, $z^\prime$	&	2008.87	\\
J21010793-4158536 &		&	M0.0	&	52	&	8.98	&	-3.46	&	$z^\prime$	&	2008.87	\\
J21073678-1304581 &		&	M3.0	&	27	&	8.73	&	-3.16	&	$z^\prime$	&	2008.87	\\
J21103147-2710578\tablefootmark{c} &		&	M4.5	&	16	&	10.30	&	-3.00	&	$i^\prime$, $z^\prime$	&	2008.87	\\
J21235271-3908176 &		&	M3.5	&	32	&	9.33	&	-3.36	&	$z^\prime$	&	2008.87	\\
J21505366-0553186 &		&	M1.0	&	51	&	9.38	&	-3.04	&	$z^\prime$	&	2008.87	\\
J21574119-5100221\tablefootmark{k} &	GJ 841 A	&	M2.5	&	16\tablefootmark{d}	&	6.75	&	-3.17	&	$i^\prime$, $z^\prime$	&	2008.87	\\
J22114208-2044181 &		&	M3.5	&	39	&	9.65	&	-3.18	&	$i^\prime$, $z^\prime$	&	2008.87	\\
J22171899-0848122\tablefootmark{c} &	GJ 852 A	&	M4.0	&	10	&	9.02	&	-2.75	&	$i^\prime$, $z^\prime$	&	2008.87	\\
J22174316-1546452 &		&	M4.0	&	22	&	10.79	&	-2.83	&	$z^\prime$	&	2008.87	\\
J22184009-5326405 &		&	M2.5	&	37	&	9.24	&	-3.25	&	$i^\prime$, $z^\prime$	&	2008.87	\\
J22230696-1736250 &	GJ 4274	&	M4.0	&	7\tablefootmark{f}	&	8.24	&	-3.26	&	$z^\prime$	&	2008.87	\\
J22332264-0936537\tablefootmark{c} &	GJ 4282	&	M2.5	&	26	&	8.53	&	-2.96	&	$i^\prime$, $z^\prime$	&	2008.87	\\
J22382974-6522423\tablefootmark{c} &	GJ 865	&	M3.5	&	15\tablefootmark{d}	&	7.27	&	-3.37	&	$i^\prime$, $z^\prime$	&	2008.87	\\
J22401867-4931045\tablefootmark{c} &		&	M5.5	&	10	&	9.84	&	-2.90	&	$i^\prime$, $z^\prime$	&	2008.87	\\
J23115362-4508004\tablefootmark{l} &	HD 218860B	&	M3.0	&	44	&	9.72	&	-2.54	&	$i^\prime$, $z^\prime$	&	2008.88	\\
J23131671-4933154 &		&	M4.0	&	15	&	9.76	&	-2.68	&	$z^\prime$	&	2008.87	\\
J23261069-7323498 &		&	M0.0	&	46	&	8.84	&	-3.09	&	$z^\prime$	&	2008.87	\\
J23285763-6802338 &		&	M2.5	&	38	&	9.26	&	-3.07	&	$i^\prime$, $z^\prime$	&	2008.87	\\
J23314492-0244395 &	GJ 1285	&	M4.5	&	11	&	9.51	&	-2.76	&	$i^\prime$, $z^\prime$	&	2008.87	\\
J23320018-3917368 &		&	M3.0	&	29	&	8.90	&	-2.83	&	$i^\prime$, $z^\prime$	&	2008.88	\\
J23323085-1215513 &		&	M0.0	&	25	&	7.45	&	-3.15	&	$z^\prime$	&	2008.87	\\
J23324655-1645081\tablefootmark{n} &	Gl 897	&	M2.5	&	12	&	6.71	&	-3.30	&	$z^\prime$	&	2008.87	\\
J23341101-1531012\tablefootmark{o} &		&	M0.0	&	47	&	8.91	&	-3.30	&	$z^\prime$	&	2008.87	\\
J23452225-7126505 &		&	M3.5	&	48	&	10.19	&	-2.88	&	$z^\prime$	&	2008.87	\\
J23474694-6517249 &		&	M1.5	&	42	&	9.10	&	-3.42	&	$z^\prime$	&	2008.87	\\
J23483610-2739385 &	GJ 4362	&	M2.5	&	27	&	8.58	&	-3.23	&	$z^\prime$	&	2008.87	\\
J23532520-7056410 &		&	M3.5	&	24	&	8.68	&	-3.53	&	$z^\prime$	&	2008.87	\\
J23555512-1321238 &	NLTT 58441	&	M2.5	&	39	&	9.26	&	-3.46	&	$z^\prime$	&	2008.87	\\
J23571934-1258406\tablefootmark{p} &	GJ 4379 B	&	M3.0	&	32	&	9.13	&	-2.88	&	$z^\prime$	&	2008.87	\\
J23572056-1258487\tablefootmark{p} &	GJ 4378 A	&	M4.0	&	23	&	8.64	&	-3.03	&	$z^\prime$	&	2008.87	\\
J23581366-1724338\tablefootmark{c} &	NLTT 58589	&	M2.0	&	28	&	8.31	&	-3.14	&	$i^\prime$, $z^\prime$	&	2008.87	\\
\end{longtable}
\tablefoot{
\tablefoottext{a}{Integrated spectral types, J magnitude, and log[$L_\mathrm{x}/L_\mathrm{bol}$] from \citet{Riaz2006}. Distance is spectroscopic distance from \citet{Riaz2006} if 
not otherwise indicated. The uncertainty in the spectroscopic distances is 37\%.}
\tablefoottext{b}{Epoch of $z^\prime$-band observations, for which the astrometric properties of the multiple systems are derived. }
\tablefoottext{c}{Binary/multiple system observed with AstraLux, see Tables 2 and 3 for properties.}
\tablefoottext{d}{Parallax distance from \textit{Hipparcos} \citep{Perryman1997}.}
\tablefoottext{e}{Parallax distance from \citet{Reid2004}.}
\tablefoottext{f}{Parallax distance from \citet{vanAltena1995}.}
\tablefoottext{g}{This is one component of a wide ($\rho=19.9\arcsec$, epoch 2001) visual double-star system \citep{Mason2001}. The system is therefore not included in 
the statistical analysis.}
\tablefoottext{h}{This is the secondary component of a wide multiple system with the G-type primary star NLTT 4704. The system is therefore not included in the
statistical analysis.}
\tablefoottext{i}{This is the primary component of a \textit{Hipparcos} visual double system, with the secondary component outside our field of view at $\rho=105.0\arcsec$ 
\citep[epoch 1991.25, ][]{Dommanget2000}. The system is not included in the statistical analysis.}
\tablefoottext{j}{The target J06061342-0337082 has spectroscopic distance 57\,pc and is therefore not included in the multiplicity analysis.}
\tablefoottext{k}{The star is part of a wide binary system with the white dwarf GJ 841 B \citep{Holberg2002} and is therefore not included in the statistical analysis.}
\tablefoottext{l}{The star is part of a wide binary system with the primary G8V star HD 218860A \citep{Torres2006} and is therefore not included in the statistical analysis.}
\tablefoottext{m}{Epoch refers to the $i^\prime$-band observation, from which astrometric properties were derived.}
\tablefoottext{n}{The star is part of a multiple system in which the K6V star \citep{Torres2006} Gl 898 is the primary \citep{Dommanget2000}. The star is therefore 
not included in the statistical analysis.}
\tablefoottext{o}{The star is part of a wide binary system \citep{Mason2001} and is therefore not included in the statistical analysis.}
\tablefoottext{p}{The star is part of a common proper motion system \citep{Mason2001} and is therefore not included in the statistical analysis.}
}
}

\begin{table*}
\caption{Photometric and astrometric properties of the observed binary/multiple M dwarfs.}
\centering
\begin{tabular}{lrrrrllr}
\hline\hline
Primary ID & $\Delta z^\prime$ & $\Delta i^\prime$  & $\rho$ & $\theta$ & New\ft{a} & $f_\mathrm{Mult}$\ft{b} & $q$\ft{c} \\
(2MASS)	  &		&		&[$\arcsec$]		&[$\degr$] & (y/n)	&	&	\\	
\hline
J00150240-7250326		&	2.13$\pm$0.18	&	1.31$\pm$0.14		&	0.290$\pm$0.009		&	69.1 $\pm$ 0.3	&	y	&	y	&	y	\\
J00250428-3646176		&	2.99$\pm$0.21	&	2.59$\pm$0.23		&	0.605$\pm$0.012		&	242.4 $\pm$ 0.3	&	y	&	y	&	y	\\
J00503319+2449009		&	0.79$\pm$0.01	&	0.94$\pm$0.02		&	1.305$\pm$0.002		&	318.9 $\pm$ 0.3	&	n	&	y	&	y	\\
J01071194-1935359		&	1.16$\pm$0.05	&	0.65$\pm$0.03		&	0.417$\pm$0.001		&	170.0 $\pm$ 0.3	&	y 	&	y	&	y	\\
J01093874-0710497		&	1.90$\pm$0.01	&	2.02$\pm$0.01		&	2.554$\pm$0.001		&	74.5 $\pm$ 0.3	&	n	&	y	&	y	\\
J01132817-3821024		&	0.37$\pm$0.01	&	0.22$\pm$0.01		&	1.405$\pm$0.003		&	29.0 $\pm$ 0.3	&	y	&	y	&	y	\\
J01365516-0647379		&	5.07$\pm$0.01	&	...			&	5.587$\pm$0.004		&	179.9 $\pm$ 0.3	&	y	&	y	&	n	\\
J01452133-3957204\tablefootmark{d}	&	2.00$\pm$0.10	&	2.39$\pm$0.16		&	0.943$\pm$0.009		&	130.9 $\pm$ 0.3	&	y	&	n	&	n	\\
J01535076-1459503		&	0.01$\pm$0.01	&	0.13$\pm$0.01		&	2.876$\pm$0.001		&	291.9 $\pm$ 0.3	&	y	&	y	&	y	\\
J02002975-0239579		&	0.75$\pm$0.06	&	0.89$\pm$0.05		&	0.323$\pm$0.001		&	5.9 $\pm$ 0.3	&	y 	&	y	&	y	\\
J02133021-4654505		&	0.73$\pm$0.27	&	1.45$\pm$0.08		&	0.135$\pm$0.001		&	124.9 $\pm$ 0.3	&	y 	&	y	&	y	\\
J02155892-0929121AB		&	2.62$\pm$0.05	&	2.80$\pm$0.05		&	0.631$\pm$0.001		&	292.2 $\pm$ 0.3	&	y 	&	y	&	y	\\
J02155892-0929121AC		&	4.99$\pm$0.05	&	5.52$\pm$0.10		&	3.509$\pm$0.002		&	299.3 $\pm$ 0.3	&	y 	&	y	&	y	\\
J02165488-2322133		&	0.91$\pm$0.01	&	0.92$\pm$0.01		&	4.369$\pm$0.001		&	314.1 $\pm$ 0.3	&	n	&	y	&	y	\\
J02271603-2929263		&	1.67$\pm$0.02	&	1.77$\pm$0.01		&	1.939$\pm$0.001		&	236.8 $\pm$ 0.3	&	y 	&	y	&	y	\\
J02335984-1811525		&	0.29$\pm$0.01	&	-1.02$\pm$0.05		&	0.854$\pm$0.001		&	48.9 $\pm$ 0.3	&	y 	&	y	&	y	\\
J02411909-5725185		&	1.31$\pm$0.02	&	1.47$\pm$0.02		&	1.526$\pm$0.001		&	287.1 $\pm$ 0.3	&	y	&	y	&	y	\\
J02451431-4344102		&	1.12$\pm$0.07	&	0.87$\pm$0.03		&	0.257$\pm$0.001		&	214.6 $\pm$ 0.3	&	y	&	y	&	y	\\
J02490228-1029220AB		&	1.33$\pm$0.06	&	1.42$\pm$0.12		&	0.481$\pm$0.006		&	209.5 $\pm$ 0.3	&	y	&	y	&	y	\\
J02490228-1029220AC		&	1.39$\pm$0.06	&	1.49$\pm$0.11		&	0.622$\pm$0.012		&	210.7 $\pm$ 0.3	&	y	&	y	&	y	\\
J03033668-2535329		&	5.14$\pm$0.06	&	3.69$\pm$0.16		&	0.834$\pm$0.005		&	7.6 $\pm$ 0.3	&	n	&	y	&	y	\\
J03050976-3725058		&	0.93$\pm$0.10	&	0.94$\pm$0.07		&	0.242$\pm$0.004		&	53.7 $\pm$ 0.3	&	y	&	y	&	y	\\
J04071148-2918342\tablefootmark{d}	&	0.70$\pm$0.05	&	0.48$\pm$0.20		&	0.295$\pm$0.001		&	44.4 $\pm$ 0.3	&	y	&	n	&	n	\\
J04080543-2731349		&	1.78$\pm$0.06	&	1.00$\pm$0.10		&	0.181$\pm$0.005		&	218.1 $\pm$ 0.3	&	y	&	y	&	y	\\
J04132663-0139211		&	0.95$\pm$0.02	&	-0.45$\pm$0.03		&	0.771$\pm$0.001		&	358.8 $\pm$ 0.3	&	n	&	y	&	y	\\
J04373746-0229282\tablefootmark{d,f}	&	1.39$\pm$0.16	&	2.57$\pm$0.05		&	0.221$\pm$0.002		&	20.5 $\pm$ 0.3	&	n	&	n	&	n	\\
J04441107-7019247		&	0.79$\pm$0.01	&	1.09$\pm$0.01		&	2.654$\pm$0.001		&	157.5 $\pm$ 0.3	&	n	&	y	&	y	\\
J05241914-1601153		&	0.36$\pm$0.03	&	0.43$\pm$0.01		&	0.639$\pm$0.001		&	69.1 $\pm$ 0.3	&	y	&	y	&	y	\\
J05254166-0909123		&	0.45$\pm$0.07	&	0.53$\pm$0.09		&	0.616$\pm$0.004		&	58.8 $\pm$ 0.3	&	n	&	y	&	y	\\
J06161032-1320422		&	1.94$\pm$0.12	&	1.40$\pm$0.23		&	0.194$\pm$0.008		&	170.6 $\pm$ 0.3	&	y	&	y	&	y	\\
J06583980-2021526AB		&	0.25$\pm$0.01	&	0.33$\pm$0.01		&	1.420$\pm$0.001		&	199.0 $\pm$ 0.3	&	y	&	y	&	y	\\
J06583980-2021526AC\tablefootmark{e}	&	5.89$\pm$0.02	&	...	&	6.992$\pm$0.002		&	263.3 $\pm$ 0.3	&	y	&	n	&	n	\\
J06583980-2021526AD		&	6.91$\pm$0.03	&	...	&	5.149$\pm$0.001		&	253.9 $\pm$ 0.3	&	y	&	y	&	n	\\
J07102991-1637350AB		&	0.46$\pm$0.09	&	0.64$\pm$0.05		&	0.568$\pm$0.001		&	354.9 $\pm$ 0.3	&	y	&	y	&	y	\\
J07102991-1637350AC		&	5.26$\pm$0.46	&	...	&	2.021$\pm$0.008		&	287.6 $\pm$ 0.3	&	y	&	y	&	y	\\
J07105990-5632596		&	1.83$\pm$0.07	&	4.17$\pm$0.07		&	1.120$\pm$0.006		&	309.8 $\pm$ 0.3	&	y	&	y	&	y	\\
J07174710-2558554\tablefootmark{f}	&	6.35$\pm$0.04	&	...	&	5.332$\pm$0.002		&	126.8 $\pm$ 0.3	&	y	&	n	&	n	\\
J07285137-3014490		&	1.29$\pm$0.11	&	1.50$\pm$0.18		&	0.485$\pm$0.002		&	169.9 $\pm$ 0.3	&	n	&	y	&	y	\\
J08224744-5726530AB		&	4.47$\pm$0.04	&	5.32$\pm$0.05		&	0.648$\pm$0.002		&	128.7 $\pm$ 0.3	&	y	&	y	&	n	\\
J08224744-5726530AC\tablefootmark{e}	&	1.83$\pm$0.04	&	...	&	8.429$\pm$0.001		&	26.1 $\pm$ 0.3	&	n	&	n	&	n	\\
J19425324-4406278		&	...	&	1.39$\pm$0.11		&	0.836$\pm$0.002		&	349.8 $\pm$ 0.3	&	y	&	y	&	y	\\
J19432464-3722108		&	2.84$\pm$0.08	&	2.89$\pm$0.07		&	1.623$\pm$0.004		&	303.7 $\pm$ 0.3	&	y	&	y	&	y	\\
J20100002-2801410		&	0.80$\pm$0.04	&	0.75$\pm$0.03		&	0.615$\pm$0.001		&	280.4 $\pm$ 0.3	&	y	&	y	&	y	\\
J20500010-1154092		&	1.04$\pm$0.22	&	1.17$\pm$0.20		&	0.486$\pm$0.046		&	348.3 $\pm$ 0.3	&	y	&	y	&	y	\\
J21103147-2710578\tablefootmark{e}	&	1.07$\pm$0.01	&	1.20$\pm$0.01		&	9.501$\pm$0.003		&	313.2 $\pm$ 0.3	&	n\tablefootmark{g}	&	n	&	n	\\
J22171899-0848122AB\tablefootmark{e}	&	0.62$\pm$0.03	&	0.74$\pm$0.01		&	7.954$\pm$0.001		&	213.2 $\pm$ 0.3	&	n	&	n	&	n	\\
J22171899-0848122AC\tablefootmark{e}	&	3.77$\pm$0.03	&	...	&	7.794$\pm$0.003		&	220.1 $\pm$ 0.3	&	n	&	n	&	n	\\
J22332264-0936537		&	0.12$\pm$0.01	&	0.65$\pm$0.05		&	1.421$\pm$0.028		&	98.6 $\pm$ 0.3 	&	n	&	y	&	y	\\
J22382974-6522423		&	0.23$\pm$0.01	&	0.20$\pm$0.02		&	0.842$\pm$0.001		&	155.8 $\pm$ 0.3 	&	n	&	y	&	y	\\
J22401867-4931045		&	0.14$\pm$0.01	&	0.16$\pm$0.01		&	4.039$\pm$0.001		&	41.0 $\pm$ 0.3 	&	n	&	y	&	y	\\
J23581366-1724338		&	0.01$\pm$0.01	&	0.03$\pm$0.01		&	1.989$\pm$0.001		&	355.7 $\pm$ 0.3	&	n	&	y	&	y	\\
\hline
\end{tabular}
\tablefoot{
\tablefoottext{a}{Companion discovered in this survey (y) or previously known (n).}
\tablefoottext{b}{Included in multiplicity fraction analysis (y/n).}
\tablefoottext{c}{Included in mass-ratio analysis (y/n).}
\tablefoottext{d}{The primary star spectral type is earlier than M0 (see Table 3). The system is therefore not included in the statistical analysis.}
\tablefoottext{e}{The survey is not complete for component separations greater than 6\arcsec~and these stars are therefore not included in the statistical analysis.}
\tablefoottext{f}{Our primary star is the secondary star in a known binary system in which the primary star is of spectral type F. The system is therefore not included in the
statistical analysis.}
\tablefoottext{g}{The companion is the star 2MASS J21103096-2710513. Although the position of the secondary star is previously known, we could find no references to the couple as
a common proper motion pair.}
}
\end{table*}

\begin{table*}
\caption{Individual spectral types and projected separations.}
\centering
\begin{tabular}{lrrr}
\hline\hline
2MASS ID & Primary  & Secondary  & Separation  \\
  & SpT	& SpT	& [A.U.] \\
\hline
J00150240-7250326	&	M0.5	&	M3.5	&	11.0	$\pm$	4.1	\\
J00250428-3646176	&	M2.5	&	M5.0	&	16.9	$\pm$	6.3	\\
J00503319+2449009	&	M3.5	&	M4.5	&	15.7	$\pm$	1.3	\\
J01071194-1935359	&	M0.5	&	M2.5	&	12.5	$\pm$	4.6	\\
J01093874-0710497	&	M1.0	&	M4.0	&	97.1	$\pm$	13.0	\\
J01132817-3821024	&	M0.0	&	M1.0	&	52.0	$\pm$	19.2	\\
J01365516-0647379	&	M4.0	&	$>$L0	&	206.7	$\pm$	76.5	\\
J01452133-3957204\tablefootmark{a}	&	K7.5	&	M3.5	&	30.2	$\pm$	2.2	\\
J01535076-1459503	&	M3.0	&	M3.0	&	51.8	$\pm$	19.2	\\
J02002975-0239579	&	M3.5	&	M4.5	&	15.5	$\pm$	5.7	\\
J02133021-4654505	&	M4.0	&	M5.0	&	1.8	$\pm$	0.7	\\
J02155892-0929121AB	&	M2.5	&	M5.0	&	16.4	$\pm$	6.1	\\
J02155892-0929121AC	&	M2.5	&	M8.0	&	91.2	$\pm$	33.8	\\
J02165488-2322133	&	M3.5	&	M4.5	&	174.8	$\pm$	64.7	\\
J02271603-2929263	&	M3.5	&	M5.0	&	98.9	$\pm$	36.6	\\
J02335984-1811525	&	M3.0	&	M3.5	&	42.7	$\pm$	15.8	\\
J02411909-5725185	&	M2.5	&	M4.0	&	68.7	$\pm$	25.4	\\
J02451431-4344102	&	M4.0	&	M4.5	&	1.8	$\pm$	0.7	\\
J02490228-1029220AB	&	M1.5	&	M3.5	&	16.4	$\pm$	6.1	\\
J02490228-1029220AC	&	M1.5	&	M3.5	&	21.2	$\pm$	7.8	\\
J03033668-2535329	&	M0.0	&	M6.0	&	32.5	$\pm$	2.8	\\
J03050976-3725058	&	M1.5	&	M3.0	&	10.9	$\pm$	4.0	\\
J04071148-2918342\tablefootmark{a}	&	K7.5	&	M1.0	&	15.1	$\pm$	5.6	\\
J04080543-2731349	&	M3.5	&	M4.5	&	7.8	$\pm$	2.9	\\
J04132663-0139211	&	M4.0	&	M4.0	&	9.3	$\pm$	3.4	\\
J04373746-0229282\tablefootmark{a,c}	&	K7.5	&	M3.0	&	5.1	$\pm$	1.9	\\
J04441107-7019247	&	M1.0	&	M2.5	&	50.4	$\pm$	18.7	\\
J05241914-1601153	&	M4.5	&	M5.0	&	5.1	$\pm$	1.9	\\
J05254166-0909123	&	M3.5	&	M4.0	&	13.6	$\pm$	5.0	\\
J06161032-1320422	&	M3.5	&	M5.0	&	6.0	$\pm$	2.2	\\
J06583980-2021526AB	&	M4.0	&	M4.0	&	45.4	$\pm$	16.8	\\
J06583980-2021526AC\tablefootmark{b}	&	M4.0	&	$>$L0	&	223.7	$\pm$	82.8	\\
J06583980-2021526AD	&	M4.0	&	$>$L0	&	164.8	$\pm$	61.0	\\
J07102991-1637350AB	&	M2.5	&	M3.0	&	27.8	$\pm$	10.3	\\
J07102991-1637350AC	&	M2.5	&	M9.0	&	99.0	$\pm$	36.6	\\
J07105990-5632596	&	M1.5	&	M4.5	&	58.2	$\pm$	21.6	\\
J07174710-2558554\tablefootmark{c}	&	M2.0	&	$>$L0	&	250.6	$\pm$	92.7	\\
J07285137-3014490	&	M1.0	&	M3.0	&	7.8	$\pm$	0.3	\\
J08224744-5726530AB	&	M4.5	&	$>$L0	&	5.2	$\pm$	1.9	\\
J08224744-5726530AC\tablefootmark{b}	&	M4.5	&	M6.0	&	67.4	$\pm$	25.0	\\
J19425324-4406278	&	M3.5	&	M4.5	&	30.1	$\pm$	11.1	\\
J19432464-3722108	&	M3.5	&	M6.0	&	50.3	$\pm$	18.6	\\
J20100002-2801410	&	M2.5	&	M3.5	&	16.0	$\pm$	5.9	\\
J20500010-1154092	&	M3.5	&	M4.5	&	18.5	$\pm$	7.1	\\
J21103147-2710578\tablefootmark{b}	&	M4.5	&	M5.5	&	152.0	$\pm$	56.3	\\
J22171899-0848122AB\tablefootmark{b}	&	M4.0	&	M4.5	&	79.5	$\pm$	29.4	\\
J22171899-0848122AC\tablefootmark{b}	&	M4.0	&	M8.5	&	77.9	$\pm$	28.8	\\
J22332264-0936537	&	M2.5	&	M3.0	&	37.0	$\pm$	13.7	\\
J22382974-6522423	&	M3.5	&	M3.5	&	12.6	$\pm$	1.2	\\
J22401867-4931045	&	M5.5	&	M5.5	&	40.4	$\pm$	14.9	\\
J23581366-1724338	&	M2.0	&	M2.0	&	55.7	$\pm$	20.6	\\
\hline
\end{tabular}
\tablefoot{
\tablefoottext{a}{The primary star spectral type is earlier than M0 (see also Appendix). The system is therefore not included in the statistical analysis.}
\tablefoottext{b}{The survey is not complete for component separations greater than 6\arcsec and these stars are therefore not included in the statistical analysis (see Table 2).}
\tablefoottext{c}{Our primary star is the secondary star in a known binary system in which the primary star is of spectral type F. The system is therefore not included in the
statistical analysis.}
}
\end{table*}

\begin{table*}
\caption{Number of systems used in multiplicity and mass-ratio analysis.}
\centering
\begin{tabular}{l r r r r}
\hline\hline
Fraction & Single & Binary & Triple & Quadruple\\
\hline

$f_\mathrm{Mult}$	&	70	&	34	&	4	&	0 \\
$q$			&	...	&	33	&	3	&	0 \\

\hline
\end{tabular}
\end{table*}

\begin{table*}
\caption{Separations and position angles for previously known multiple systems.}
\centering
\begin{tabular}{l r r r r}
 \hline\hline
2MASS ID & $\rho$ & $\theta$ & Epoch & Ref.\\
  &	[$\arcsec$] & 	 [$\degr$]	&	&	\\
\hline
	
J00503319+2449009	&	1.0	&	315	&	1960	&	1	\\
	&	2.080	&	316.0	&	1991.25	&	2	\\
	&	1.648	&	317.12	&	2002.64	&	3	\\
	&	1.305	&	318.9	&	2008.86	&	4	\\
\\
J01093874-0710497	&	2.680	&	77.0	&	1991.25	&	2	\\
	&	2.554	&	74.5	&	2008.87	&	4	\\
\\
J02165488-2322133	&	4.3	&	315	&	1998.67	&	5	\\
	&	4.369	&	314.1	&	2008.87	&	4	\\
\\
J04132663-0139211	&	0.79	&	217.11	&	1998.9	&	6	\\
	&	0.771	&	358.8/178.8	&	2008.88	&	4	\\
\\
J04373746-0229282	&	0.225	&	195	&	2003.05	&	7	\\
	&	0.093	&	189.5	&	2004.95	&	7	\\
	&	0.221	&	20.5	&	2008.88	&	4	\\
\\
J04441107-7019247	&	2.3	&	174	&	1990	&	8	\\
	&	2.654	&	157.5	&	2008.88	&	4	\\
\\
J05254166-0909123	&	0.537	&	69.40	&	2005.78	&	9	\\
	&	0.616	&	58.8	&	2008.86	&	4	\\
\\
J07285137-3014490	&	0.175	&	143.71	&	2002.83	&	9	\\
	&	0.485	&	169.9	&	2008.86	&	4	\\
\\
J08224744-5726530	&	8.6	&	23	&	1999.99	&	5	\\
	&	8.429	&	26.1	&	2008.88	&	4	\\
\\
J21103147-2710578	&	9.4	&	313	&	1998.59	&	5	\\
	&	9.501	&	313.2	&	2008.87	&	4	\\
\\
J22171899-0848122AB	&	7.8	&	213	&	1998.79	&	5	\\
	&	7.95	&	213.2	&	2008.87	&	4	\\
\\
J22171899-0848122BC	&	0.978	&	305.8	&	2001.60	&	10	\\
	&	0.97	&	316.7	&	2008.87	&	4	\\
\\
J22332264-0936537	&	1.66	&	272.25	&	1997.6	&	6	\\
	&	1.571	&	279.73	&	2005.44	&	9	\\
	&	1.421	&	98.6/278.6	&	2008.87	&	4	\\
\\
J22382974-6522423	&	0.770	&	16	&	1991.25	&	2	\\
	&	0.842	&	155.8	&	2008.87	&	4	\\
\\
J22401867-4931045	&	4.2	&	40	&	1999.72	&	5	\\
	&	4.039	&	41.0	&	2008.87	&	4	\\
\\
J23581366-1724338	&	1.904	&	355.3	&	2005.54	&	9	\\
	&	1.989	&	355.7	&	2008.87	&	4	\\

\hline

\end{tabular}
\tablebib{(1)~\citet{Dommanget2002}; (2) \citet{Perryman1997}; (3)  \citet{StrigachevLampens2004}; (4) This work;
(5) \citet{Cutri2003}; (6) \citet{McCarthy2001}; (7)  \citet{Kasper2007}; (8) \citet{Mason2001}; (9) \citet{Daemgen2007};
(10) \citet{Beuzit2004}.
}
\end{table*}

\appendix
\section{Notes on individual binaries and multiple systems}
Table 5 summarises our measured angular separations and position angles, and published results for previously known components.

\textbf{J00503319+2449009} 
This star, also known as GJ 3060A or NLTT 2805, is a flare star \citep{Norton2007} with a known stellar companion, NLTT 2804. The Catalog of Components of
Double \& Multiple Stars \citep[CCDM,][]{Dommanget2002} provides the astrometric measurements $\rho=1.0\arcsec$ and $\theta=315\degr$ for epoch 1960.
\textit{Hipparcos} observations provide positions of the two components of $\rho=2.080\arcsec$ and $\theta=316\degr$ \citep[epoch 1991.25, ][]{Perryman1997}.
\citet{StrigachevLampens2004} present photometric and astrometric observations of visual double stars and for this binary estimate the angular separation to be
$\rho = 1.648\arcsec$ and the position angle $\theta=317.12\degr$ (epoch 2002.64).
Our measured separation is $\rho=1.305\arcsec$ and position angle $\theta=318.9\degr$, indicating orbital motion.
 
\textbf{J01093874-0710497} This is a high proper motion star with $\mu_{\rm RA}$=-235.5 mas/yr and  $\mu_{\rm DEC}$=-351.6 mas/yr. Also known as HIP 5443, it is a 
\textit{Hipparcos} double star \citep{Perryman1997} with separation
$\rho=2.7\arcsec$ and position angle $\theta=77\degr$ (epoch 1991.25).
We measure the separation $\rho=2.554\arcsec$ and position angle $\theta=74.5\degr$, hence both components form a common proper motion pair. The small change in separation
and position angle in the more than 15 years that have passed between the \textit{Hipparcos} and our measurements can be attributed to orbital motion.

\textbf{J01365516-0647379} The primary star is a high proper motion star for which \citet{Shkolnik2009} estimated an age between 25 and 300 Myr. Because of its
faint magnitude, the secondary star could not be seen at the time of observation but only after additional analysis. The star therefore ended up partly outside the
field of view in the $i^\prime$-band observation and we present only $z^\prime$-band data in this paper.

\textbf{J02165488-2322133} In 2MASS PSC \citep{Cutri2003}, we find the star J02165465-2322103 at separation $\rho=4.3\arcsec$ and position angle $\theta=315\degr$ from our 
primary (epoch 1998.67), which corresponds well to our measured separation $\rho=4.369\arcsec$ and position angle $\theta=314.1\degr$. 

\textbf{J02335984-1811525} In this double system, the B component is brighter than the A component in $i^\prime$-band. Since the $i^\prime$ and $z^\prime$ band observations 
were performed on different 
nights, the unusual $i^\prime-z^\prime$ colour might indicate that the star is variable or possibly of T Tauri-type. 
We tentatively assign spectral types M3$\pm$1 + M3.5$\pm$1 to the stars, but further investigation of this couple is necessary to determine their characteristics. 

\textbf{J02490228-1029220} The B and C components of this triple star are close, $\rho_\mathrm{BC}=0.145 \arcsec$ corresponding to 4.93\,AU.

\textbf{J03033668-2535329} The primary star is a high proper motion star also known as  NLTT 9775.
We measure a separation $\rho=0.834$\arcsec~between the two companions. 
A possible candidate for the secondary star is the high proper motion star LTT 1453, which has J2000 coordinates RA=03h 03m 36.6s, 
Dec= -25\degr 35\arcmin 33\arcsec, at an angular separation of 1.42\arcsec from our primary star.
\citet{Frankowski2007} studied the binary content of the
\textit{Hipparcos} catalogue, listing the primary star as a candidate proper motion binary.

\textbf{J04080543-2731349} The images in both $i^\prime$- and $z^\prime$-band of this binary are affected by ``fake tripling''. The real B component and the fake triple are 
equally bright in $z^\prime$-band but unequal in $i^\prime$. This means that, although unlikely, the true position angle might be systematically incorrect by 180\degr. 

\textbf{J04132663-0139211} This binary system was discovered by \citet{McCarthy2001}, with a separation $\rho=0.79\arcsec$ and position angle
$\theta=217.11\degr$ (epoch 1998.9). We measure $\rho=0.771\arcsec$ and $\theta=358.8\degr$, indicating significant orbital motion between observations. In our
observations,
the B component is brighter than the A component in $i^\prime$-band. Since the $i^\prime$ and $z^\prime$ band observations were performed on different nights, 
the unusual $i^\prime-z^\prime$ colour may indicate that the star is variable or possibly of T Tauri-type. If our secondary star is the primary star of 
\citet{McCarthy2001}, the position angle is instead $\rho=178.8\degr$. We tentatively assign the stars spectral types M$4\pm1$ + M$4\pm1$, 
but further investigation of this double system is needed to determine its character.

\textbf{J04373746-0229282} The primary star is also known as GJ 3305, a member of the young $\beta$ Pictoris moving group \citep{Zuckerman2001}, which has an 
estimated age of 12 Myr \citep{Shkolnik2009}. The faint close companion that we see was discovered by \citet{Kasper2007} in their L-band NACO imaging of young, nearby stars in 
search of substellar companions. \citet{Kasper2007} present NACO K-band data from the ESO/ST-ECF Science Archive with which they determine the separation $\rho=0.225\arcsec$ 
and position angle $\theta=195\degr$ (epoch 2003.05), and their obtained L-band data for which $\rho=0.093\arcsec$ and position angle
$\theta=189.5\degr$ (epoch 2004.95), and the proper motion combined points to a bound companion in a highly eccentric orbit. Our observations 
are affected by the stellar companion ghost image at 180$\degr$ discussed in Sect. 2.2, which may cause uncertainty in the true position angle. However, the
assumed position at 
$\rho=0.221\arcsec$ and $\theta=20.5\degr$ is consistent with physical companionship, and with the non-detection of the secondary companion by 
\citet{Daemgen2007} (epoch 2005.74) indicates that the orbit has a high
inclination, i.e., is seen close to edge-on.

\citet{Feigelson2006} agree with \citet{Zuckerman2001} and conclude from the proper 
motion and stellar activity that GJ 3305 is part of a wide binary system ($\rho=66\arcsec$, or $\sim2000$\,AU at 30\,pc) with the F0 star 51 Eri. Since the
primary star is then of earlier spectral type than M0, the system is not included in our statistical analysis.

\textbf{J04441107-7019247} This is a previously known visual binary, also known as HD 270712. \citet{Mason2001} provides the astrometric measurements 
$\rho=2.3\arcsec$ and $\theta=174\degr$ for epoch 1990. We measure $\rho=2.654\arcsec$ and $\theta=157.5\degr$, indicating orbital motion.

\textbf{J05254166-0909123} This high proper motion binary (NLTT 15049) was discovered by \citet{Daemgen2007}, at a separation of $\rho=0.537\arcsec$ and 
position angle 
$\theta=69.40\degr$ (epoch 2005.78). We assign spectral types M3.5+M4 to the couple, consistent with the spectral types of \citet{Daemgen2007}. We find a separation 
$\rho=0.616\arcsec$ and position angle $\theta=58.8\degr$, indicating
significant orbital motion. \citet{Shkolnik2009} estimates the age of the stars to between 35 and 300\,Myr. 

\textbf{J06583980-2021526} This possibly quadruple system consists of one close M4+M4 pair and two more distant suspected L dwarfs. The two faintest components (C and D)
 are separated from the brightest star by $5.15\arcsec$ and $6.99\arcsec$, respectively, and had not been discovered at the time of observation. They therefore ended up outside 
the FoV in $i^\prime$-band. The separation between components A and B is $\rho_\mathrm{AB}=1.420\arcsec$, and the two faint stars C and D are separated by
$\rho_\mathrm{CD}=2.093\arcsec$ with position angle $\theta_\mathrm{CD}=107.3\degr$. The separation between the primary star and the C component is greater than our completeness 
limit, and the component is therefore not included in the statistical analysis.

\textbf{J07102991-1637350} The tertiary companion is too faint in $i^\prime$ for accurate photometry and astrometry. We therefore present only $z^\prime$-band data in this 
paper.

\textbf{J07105990-5632596} We obtain different spectral types for the secondary star in $i^\prime$ and $z^\prime$ (SpT$_{i^\prime}\approx$ M5.5, SpT$_{z^\prime}\approx$ M4). 
Since we did not observe the stars in both filters on the same night, the brightness of 
either companion might have changed from one observation to the next if the stars are variable. We tentatively assign the secondary spectral type M4.5$\pm$1.

\textbf{J07174710-2558554} Our primary star is also known as CD-25 4322B, the secondary star in a wide double system with CD-25 4322. Our wide but faint
secondary component is not, however, found in any catalogue. The star CD-25 4322 is an F-star \citep[F0/F3V, ][]{Dommanget2002, Perryman1997} and not 
within our field of view \citep[CCDM separation $\rho=12.4\arcsec$, epoch 1897, ][]{Dommanget2002}. 
Because of the faintness of our secondary companion, it was not detected at the time of observation and not observed in $i^\prime$. Since our primary star is 
the secondary star in a system with an F-star primary, it is not included in the statistical analysis.

\textbf{J07285137-3014490} This is a known binary system also known as GJ 2060 \citep{Zuckerman2004}, which was concluded by \citet{AllenReid2008} to probably be part
of a quadruple system with another close, equal mass M dwarf binary at a separation of $\rho=67.2\arcsec$. GJ 2060 is a likely member of the $\sim$50\,Myr old AB Dor 
association \citep{Zuckerman2004}.
We obtain spectral types M1+M3, while \citet{Daemgen2007} find
spectral types M0.5+M1.5. The primary star is a known variable star (V372 Pup).
\citet{Daemgen2007} determine the binary separation and position angle to be $\rho=0.175\arcsec$ and $\theta=143.71\degr$ (epoch 2002.83), while we find 
$\rho=0.485\arcsec$ and $\theta=169.9\degr$, indicating significant orbital motion.

\textbf{J08224744-5726530} The primary star of this triple system is also known as LHS 2005, a high proper motion star. 
Our separation and position angle for the tertiary component, which is also known as LHS 2004, is 
$\rho=8.429\arcsec$ and $\theta=26.1\degr$. This is consistent with data from the 2MASS PSC \citep{Cutri2003} for the star J08224787-5726451 with a separation 
$8.6\arcsec$ and 
position angle $\theta=23\degr$ (epoch 1999.99) and is indicative of orbital motion. LHS 2004 and LHS 2005 form a known common proper motion pair. The close secondary was not 
previously known. 
The wide companion was noticed at the time of observation and fitted into the field of view by placing the primary star in the corner of the detector 
for the $z^\prime$-band observations. The wide companion is outside the field of view in $i^\prime$-band. Since the C component separation from the primary is greater than 
6\arcsec,~it is not included in the statistical analysis. 

\textbf{J19425324-4406278} The secondary star is previously unknown. Only $i^\prime$-band images could be used in our analysis since the secondary star was too faint in 
$z^\prime$. The position angle, separation, and individual spectral types are therefore obtained from the $i^\prime$-band observation.

\textbf{J21103147-2710578} The companion is J21103096-2710513 at a 2MASS PSC distance of $9.4\arcsec$ and position angle $\theta=313\degr$\citep[epoch 1998.59, ][]
{Cutri2003}. 
We find a separation $\rho=9.50\arcsec$, which is greater than our limits for completeness. The system is therefore not included in the statistical analysis. 

\textbf{J22171899-0848122} This is a known visual binary system where the primary star (also known as V* FG Aqr or GJ 852A) and the secondary star 
(J22171870-0848186, or GJ 852B) are both flare stars \citep{Gershberg1999}.
The tertiary companion, close to our secondary star GJ 852B at $\rho_\mathrm{BC}=0.97\arcsec$ and $\theta_\mathrm{BC}=316.7\degr$, was discovered by \citet{Beuzit2004} 
at $\rho=0.978\arcsec$ and $\theta=305.8\degr$ (epoch 2001.60), hence the system shows orbital motion. The C component is in our observations too faint to 
be resolved in $i^\prime$ band. Photometric measurements in $i^\prime$ for the B component therefore include the very faint flux from the close C component.
The 2MASS PSC \citep{Cutri2003} infer a proximity of $7.8\arcsec$ and position angle of $\theta=213\degr$ (epoch 1998.79), relating the positions of GJ 852 A 
and GJ 852 B. Our measured separation between these stars is $\rho=7.95\arcsec$ and position angle $\theta=213.2\degr$. 

\textbf{J22332264-0936537} Also known as GJ 4282, this flare star was discovered to be a binary by \citet{McCarthy2001}, who derived a separation of $\rho=1.66\arcsec$ and 
position angle $\theta=272.25\degr$ for epoch 1997.6. \citet{Daemgen2007} observed a separation of $\rho=1.571\arcsec$ and $\theta=279.73\degr$
for epoch 2005.44.
We find $\rho=1.421\arcsec$ and $\theta=98.6\degr$, a separation that agrees with previous observations but at a position angle that is clearly inconsistent with
the previous measurements by \citet{Daemgen2007} and \citet{McCarthy2001}. With an estimated orbital period of approximately 380 years, we need to assume that our primary 
star (the eastern component) is actually the secondary star of \citet{Daemgen2007} and \citet{McCarthy2001}, and our revised position angle is in that case 
$\theta=278.6\degr$,
indicating orbital motion. 
Since in our observations the eastern star is slightly brighter than the western component, one or both of the stars might be variable, causing
the discrepancy in position angle between our observations and the observations by \citet{McCarthy2001} and \citet{Daemgen2007}.
We assign the stars spectral types M2.5 and M3, respectively, in agreement with \citet{Daemgen2007}(M3+M3) and \citet{Shkolnik2009} (eastern component M2.5, western component
 M2.6). 
\citet{Shkolnik2009} estimate the age of the system to be 20-150\,Myr.

\textbf{J22382974-6522423} This flare star, which is also known as GJ 865, was identified by \citet{Montes2001} as a possible member of the $\sim600$\,Myr Hyades supercluster.
The star GJ 865 is part of a known triple system. We observed the two close components, separated by $\rho=0.842\arcsec$, which is in agreement with the
separation $\rho=0.770\arcsec$ and position angle $\theta=16\degr$ found by \citet{Perryman1997}
for epoch 1991.25. The third companion is outside our field of view, with a separation from our primary star of $\rho=30.4\arcsec$
\citep[epoch 1974, ][]{Dommanget2002}. While we could not find the spectral type of this companion in literature, the V magnitudes of the three companions differ only slightly 
\citep[$V_\mathrm{A}=12.0, V_\mathrm{B}=12.1, V_\mathrm{C}=12.3$, ][where the close components are B and C]{Dommanget2002} and we assume that the third component is also an M star. We therefore
include this system in the binary statistics as an M dwarf binary/multiple system.  

\textbf{J22401867-4931045} This couple of high proper motion stars \citep{Lepine2005} are also known as LSR J22403-4931W (our primary star) and LSR J22403-4931E located at 
RA = 22h\,40m\,18.96s, Dec = -49\degr31\arcmin01.4\arcsec (J2000). \citet{Cutri2003} found $\rho=4.2\arcsec$ and $\theta=40\degr$ for epoch 1999.72. We measure 
$\rho=4.039\arcsec$ and $\theta=41.0\degr$.

\textbf{J23581366-1724338} The binary character of this high proper motion star, also known as NLTT 58589, was discovered by \citet{Daemgen2007}, who derived the same 
individual spectral types M2+M2, as we do.
We find $\rho=1.989\arcsec$, in good agreement with the \citet{Daemgen2007} separation $\rho=1.904\arcsec$ for epoch 2005.54, although our measured position 
angle $\theta=355.7\degr$ disagrees with the \citet{Daemgen2007} result of $\theta=265.30\degr$ by $90\degr$. Reanalysis of the Gemini data by Daemgen et al. yields a 
position angle of $355.3\degr$, which is in good agreement with the AstraLux Sur measurement and indicates some orbital motion.
\citet{Shkolnik2009} determine individual spectral types M1.9 (north)+M1.9 (south) and an age of 20-150\,Myr for the system.

\textbf{J23534173-6556543}: We also observed this star, which is the secondary star in a widely separated G0\,V+M1\,V system, and its primary. 
The primary is HIP 117815 and the secondary is CPD-66 3810B.
Our separation of $\rho=12.3\arcsec$ at $\theta=112.2\degr$ is in good agreement with \citet{Eggenberger2007} 
($\rho=12.14\arcsec, \theta=112.37\degr$, epoch 2005.70) for this bound system. This system was only observed in $z^\prime$-band and is not included in any statistical 
analysis in this paper because the primary star is a G star.

\end{document}